\documentclass{article}
\pdfoutput=1
\usepackage{PRIMEarxiv}
\usepackage[utf8]{inputenc} 
\usepackage[T1]{fontenc}    
\usepackage{hyperref}       
\usepackage{url}            
\usepackage{booktabs}       
\usepackage{nicefrac}       
\usepackage{microtype}      
\usepackage{lipsum}
\usepackage{fancyhdr}       
\usepackage{amsmath,amsfonts,amssymb,xcolor,mathtools}
\usepackage{mathrsfs}
\usepackage{graphicx} 
\usepackage{epstopdf,epsfig}
\usepackage{multirow}

\newcommand{\R}{\mathbb{R}}

\newcommand{\N}{\mathbb{N}}
\newcommand{\Z}{\mathbb{Z}}

\newcommand{\upi}{\mathrm{i}}
\renewcommand{\Re}{\mathrm{Re}}
\renewcommand{\Im}{\mathrm{Im}}
\DeclareMathAlphabet{\mathsfbi}{OT1}{\sfdefault}{bx}{sl}

\newcommand{\upd}{\mathrm{d}}

\title{Water wave interactions with surface-piercing vertical barriers in a rectangular tank: Connections with Bloch waves and quasimodes}

\author{
  Ben Wilks\\
  Department of Mathematics and Statistics\\
  University of Otago\\
  Dunedin\\
  New Zealand\\
\texttt{ben.wilks@newcastle.edu.au}\\
  \AND
  Fabien Montiel\\
  Department of Mathematics and Statistics\\
  University of Otago\\
  Dunedin\\
  New Zealand\\
  \AND
  Luke G. Bennetts\\
  School of Computer and Mathematical Sciences\\
  University of Adelaide\\ Adelaide, SA 5005\\
  Australia\\
  \AND
  Sarah Wakes\\
  Department of Mathematics and Statistics\\
  University of Otago\\
  Dunedin\\
  New Zealand\\
}

\begin{document}
\maketitle
\begin{abstract}
Eigenmodes are studied for a fluid-filled rectangular tank containing one or more vertical barriers, and on which either Dirichlet or Neumann boundary conditions are prescribed on the lateral walls. In the case where the tank contains a single barrier, the geometry of the tank is equivalent to the unit cell of the cognate periodic array, and its eigenmodes are equivalent to standing Bloch waves. As the submergence depth of the barrier increases, 
it is shown that the passbands (i.e.\ frequency intervals in which the periodic array supports Bloch waves) become thinner, and that  this effect becomes stronger at higher frequencies.
The eigenmodes of a uniform array of vertical barriers in a rectangular tank are also considered. They are found to be a superposition of left- and right-propagating Bloch waves, which couple together at the lateral walls of the tank. A homotopy procedure is used to relate the eigenmodes to the quasimodes of the same uniform array in a fluid of infinite horizontal extent, and the quasimodes are shown to govern the response of the array to incident waves. Qualitative features of the mode shapes are typically preserved by the homotopy, which suggests that the resonant responses of the array in an infinite fluid can be understood in terms of modes of the array in a finite tank.
\end{abstract}
\section{Introduction}
Water wave metamaterials, which are offshore structures composed of subwavelength resonators, have been the focus of both theoretical \cite{Hu2004,Hu2011,Dupont2017,Bennetts2018,wilks2022rainbow,huang2023Water,wilks_montiel_wakes_2023} and experimental \cite{hu2013experimental,Archer2020,euve2021control,euve2023negative} studies for their promising applications, including wave energy capture and the protection of coasts and offshore structures. One popular example is an array of surface-piercing vertical barriers immersed in water. Such structures have been proposed for controlling the direction of wave propagation \cite{zheng2020wave,PORTER2021102673,zheng2022}, as well as for their implications for wave energy capture \cite{wilks2022rainbow,huang2023surface,zheng2024wave}.

Water wave propagation through arrays of vertical barriers, which underlies these promising applications, has been studied in the two dimensional (one horizontal and one vertical) linear water wave context \cite{wilks2022rainbow,huang2023Water,wilks_montiel_wakes_2023}. These studies build on earlier, foundational studies of water wave scattering by one or two vertical barriers \cite{Ursell1947,Evans1972,Newman1974,Porter1995a,McIver1985}. Arrays of vertical barriers, in which the barriers have a spatially-graded submergence depth, were shown to exhibit the rainbow reflection effect, by which broadband wave pulses propagating through the array slow down and become spatially separated by frequency, before being reflected \cite{wilks2022rainbow}. The promising implications of this rainbow reflection to broadband wave-energy conversion were also explored in that paper. Later, rainbow reflection in graded arrays of vertical barriers was shown to be accurately described by the local Bloch wave approximation, which assumes that wave interactions within each unit cell (here, a region containing a single barrier in the array) can be approximated by the propagating solutions of the cognate infinite array---the so-called Bloch waves \cite{wilks_montiel_wakes_2023}. 
Wave scattering by uniform finite arrays of vertical barriers has also been studied in the close-spacing limit using asymptotic homogenisation \cite{huang2023Water}. This method is accurate at low frequencies, but breaks down near the resonant frequency associated with vertical fluid motion between the barriers, where the assumptions of the asymptotic scheme are violated. 

Here, we present a semi-analytic study of uniform arrays of vertical barriers, which allows the resonances to be investigated directly.
Our investigation draws upon the analogy between Bloch wave propagation through a periodic array of vertical barriers and Rayleigh--Bloch waves, i.e.\ localised waves which propagate along an infinite periodic line array of cylindrical scatterers with wavenumber $\beta(k)>k$ (where $k$ is the wavenumber of the background medium) and decay exponentially in both directions away from the array \cite{porter1999rayleigh}. Rayleigh--Bloch waves exist for all $k\leq k_c$, where $k_c$ is called the cutoff wavenumber. Porter and Evans \cite{porter1999rayleigh} showed that the Rayleigh--Bloch wave at the cutoff wavenumber $k_c$, which is a standing wave, is equivalent to the trapped mode of a single cylinder in a straight-walled parallel waveguide with homogeneous Neumann boundary conditions, and width equal to the cylinder spacing of the cognate periodic array. The existence of this so-called Neumann trapped mode was previously proved by \cite{callan_linton_evans_1991}. The related problem in which homogeneous Dirichlet boundary conditions are imposed on the walls of the waveguide was studied in \cite{maniar1997wave} in connection with the resonances of a finite line array of cylinders. The authors identified a Dirichlet trapped mode which, although different from the Neumann trapped mode, also corresponds to a standing wave localised to an infinite periodic line array of cylinders with spacing equal to the width of the waveguide. The starting point of this paper is the problem of a rectangular tank with either Dirichlet or Neumann walls and containing a single vertical barrier, which is analogous to the waveguide/cylinder problem. 

The outline of this paper is as follows. In \textsection\ref{single_barrier_tank_sec}, we explore the modes of a single-barrier tank having either Dirichlet or Neumann walls. We show that the eigenmodes of the tank correspond to standing Bloch waves supported by a periodic array of vertical barriers. As a consequence, the resonant frequencies of the tank delineate the edges of the passbands. In \textsection\ref{multiple_barriers_sec}, we consider rectangular tanks containing multiple vertical barriers. In particular, we show that modes of the $(N+1)$-barrier Dirichlet and Neumann tanks correspond to both standing and propagating Bloch waves. In \textsection\ref{quasimodes_sec}, we use a homotopy procedure to show how the modes of the $(N+1)$-barrier Dirichlet and Neumann tanks relate to the quasimodes (also known as complex resonances, or nearly-trapped modes) of an array of $N+1$ barriers immersed in a fluid of infinite horizontal extent. As a consequence, these modes govern the resonant responses to incident plane wave forcing. A conclusion is given in \textsection\ref{conclusion_sec}.

\section{Single barrier in a tank}\label{single_barrier_tank_sec}
The investigation in this paper was initiated by a discrepancy between numerical results presented in \cite{wilks2022rainbow} and those computed independently using COMSOL Multiphysics(R)\footnote{The COMSOL computations were carried out by G. J. Chaplain (personal communication, May 17, 2023).}. In particular, the band diagrams (i.e., curves showing the dispersion relation of Bloch waves in a periodic array of vertical barriers) were not in complete agreement. Although both methods agreed with regards to the lowest frequency passband, COMSOL also detected a sequence of higher frequency passbands that were not reported in \cite{wilks2022rainbow}. In response, we set out to investigate whether the periodic array of vertical barriers genuinely supports these higher passbands. To do this, we first consider the related problem of the resonant modes of a fluid-filled rectangular tank containing a single vertical barrier---a problem previously considered by \cite{evans1987resonant}. These authors showed that as submergence depth of the barrier increases, the resonant frequencies of the tank interpolate between the edge cases of the barrier being either absent or fully submerged, which have closed-form expressions. The authors of \cite{evans1987resonant} only considered the case where Neumann boundary conditions are imposed on the side walls of the tank. Here, we extend their results to consider the case where non-physical Dirichlet boundary conditions are imposed on the tank walls (motivated by the previous discussion on Rayleigh--Bloch waves; see \textsection 1) and discuss the connections with wave propagation through the cognate periodic array.

\subsection{Preliminaries}
\begin{figure}
    \centering
    \includegraphics[width=0.7\textwidth]{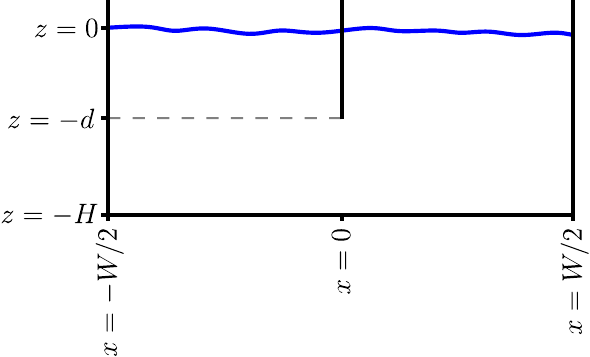}
    \caption{Schematic of the rectangular tank of width $W$ and depth $H$ containing a single surface-piercing vertical barrier of submergence depth $d$, positioned in the centre of the tank.}
    \label{fig:tank_schematic}
\end{figure}

A schematic of a tank containing a single vertical barrier is given in figure \ref{fig:tank_schematic}. At equilibrium, the fluid occupies the region $\Omega=\{(x,z)|-W/2<x<W/2,-H<z<0\}\setminus \Gamma$, where $H$ is the depth of the fluid and $x=\pm W/2$ describes the position of the tank's side walls. 
The set $\Gamma=\{(0,z)|-d<z<0\}$ describes the barrier's location, where $d$ is the submergence depth of the barrier at equilibrium.

Linear water wave theory \cite{Linton2001,Mei2005} is used to model fluid flow. Further, assuming time-harmonic motions, the velocity potential of the fluid can be expressed as $\mathrm{Re}(\phi(x,z)e^{-\upi\omega t})$, where $\omega$ is the angular frequency, $t$ is time and $\phi$ satisfies the following boundary value problem:
\begin{subequations}
\label{bvp_tank_appendix}
\begin{align}
    \bigtriangleup \phi&=0&\text{for }(x,z)\in\Omega\\
    \partial_z\phi&=0&\text{for }z=-H\\
    \partial_z\phi&=\frac{\omega^2}{g}\phi&\text{for }z=0\\
    \partial_x\phi&=0&\text{for }(x,z)\in\Gamma\label{barrier_BC_tank_appendix}
\end{align}
\end{subequations}
(where $g$ is acceleration due to gravity) together with boundary conditions on the tank walls at $x=\pm W/2$ (defined below). 
The submerged tip of the barrier induces a Meixner (or corner) singularity in the velocity potential of the form \cite{Mosig2018}
\begin{equation}
\sqrt{x^2+(z+d)^2}\|\nabla \phi\|\to 0\quad\text{as}\quad\sqrt{x^2+(z+d)^2}\to 0.
\end{equation}

The boundary conditions on the tank walls can be either of Dirichlet or Neumann type, i.e.,
\begin{subequations}\label{dir_Neu_BC}
\begin{align}
    \phi(x,z)&=0&\text{for }x=\pm W/2\text{ and }z\in(-H,0)\\[4pt]
    \text{or}\quad\partial_x\phi(x,z)&=0&\text{for }x=\pm W/2\text{ and }z\in(-H,0),
\end{align}
\end{subequations}
respectively. 
The Neumann boundary condition has the physical interpretation of a no-flow boundary, whereas the Dirichlet boundary condition is non-physical in the context of water waves. Nevertheless, it has proved important in studies connecting Rayleigh-Bloch waves with trapped modes in waveguides \cite{maniar1997wave,utsunomiya_taylor_1999,porter1999rayleigh} and we will make similar arguments here.

\subsection{Solution to the eigenvalue problem}\label{sec_method}
Following \cite{wilks2022rainbow}, the general solution is obtained using separation of variables, so that
\begin{equation}\label{general_sol}
    \phi(x,z)=\begin{cases}
        \displaystyle\sum_{m=0}^\infty \Large\{A_m^{(0)}\exp(\upi k_m x)+B_m^{(0)}\exp(-\upi k_mx)\Large\}\psi_m(z)&x<0\\[4pt]
        \displaystyle\sum_{m=0}^\infty\Large\{ A_m^{(1)}\exp(\upi k_mx)+B_m^{(1)}\exp(-\upi k_mx)\Large\}\psi_m(z)&x>0,
    \end{cases}
\end{equation}
for all $(x,z)\in\Omega$. Using the convention of \cite{Linton2001}, the quantities $k_m$, which are the solutions to the dispersion relation
\begin{equation}
    k\tanh{kH}=\omega^2/g,
\end{equation}
are ordered so that $k_0\in\R^+$ and $-\upi k_m\in{((m-1)\pi/H,m\pi/H)}$ for all ${m\in\N}$. Following \cite{Porter1995a}, the vertical eigenfunctions $\psi_m$ are defined as
\begin{equation}
    \psi_m(z)= \left(\frac{\sinh(2k_m H)}{4k_m H}+\frac{1}{2}\right)^{-1/2}\cosh(k_m(z+H)).
\end{equation}

To impose the boundary conditions  \eqref{barrier_BC_tank_appendix} and \eqref{dir_Neu_BC} on the general solution \eqref{general_sol}, we first truncate the infinite sums in \eqref{general_sol} up to $m=N_{\mathrm{Sol}}$. 
The boundary conditions on the walls \eqref{dir_Neu_BC} can then be encoded by matrix-vector expressions
\begin{subequations}\label{wall_matrix_equations}
    \begin{equation}\label{Neumann_Wall_matrix_equations}
    \mathbf{B}^{(0)}=\mathsfbi{L}\mathbf{A}^{(0)}\quad\text{and}\quad
    \mathbf{A}^{(1)}=\mathsfbi{L}\mathbf{B}^{(1)},
\end{equation}
in the Neumann case, or
\begin{equation}\label{Dirichlet_Wall_matrix_equations}
    \mathbf{B}^{(0)}=- \mathsfbi{L}\mathbf{A}^{(0)}\quad\text{and}\quad
    \mathbf{A}^{(1)}=- \mathsfbi{L}\mathbf{B}^{(1)},
\end{equation}
\end{subequations}
in the Dirichlet case. In \eqref{wall_matrix_equations}, we have introduced the coefficient vectors $\mathbf{A}^{(n)}$ and $\mathbf{B}^{(n)}$, which have entries $A^{(n)}_m$ and $B^{(n)}_m$ for $n\in\{0,1\}$, respectively, as well as the diagonal phase-shift matrix $\mathsfbi{L}$ with diagonal entries $\exp(\upi k_m W)$ for $m\in\{0\}\cup\mathbb{N}$.

The boundary condition on the barrier \eqref{barrier_BC_tank_appendix} must also be considered. In particular, the coefficients in \eqref{general_sol} must be chosen so that the potential $\phi$ satisfies the no-flow condition \eqref{barrier_BC_tank_appendix} on the barrier and is continuously differentiable beneath the barrier. After truncation of the infinite sums in \textsection\ref{general_sol} at some $m=N_{\mathrm{Sol}}$, these conditions may be encoded by a $2(N_{\mathrm{Sol}}+1)$-dimensional scattering matrix, which relates the coefficients of the system via
\begin{equation}\label{S_matrix}
    \begin{bmatrix}\mathsfbi{G}&\mathsfbi{I}- \mathsfbi{G}\\
    \mathsfbi{I}- \mathsfbi{G}&\mathsfbi{G}
    \end{bmatrix}
    \begin{bmatrix}
        \mathbf{A}^{(0)}\\\ \mathbf{B}^{(1)}
    \end{bmatrix}
    =
    \begin{bmatrix}
        \mathbf{A}^{(1)}\\\ \mathbf{B}^{(0)}
    \end{bmatrix},
\end{equation}
where $\mathsfbi{I}$ is the $(N_{\mathrm{Sol}}+1)$-dimensional identity matrix. The submatrix $\mathsfbi{G}$ is obtained by solving the problem of water wave scattering by a single surface-piercing vertical barrier using the integral equation/Galerkin method of \cite{Porter1995a}.

Given that the tank has either Dirichlet or Neumann walls, and assuming that the eigenmodes are either symmetric or antisymmetric with respect to the line $x=0$ (owing to the symmetry of the tank), there are four cases to consider.

\subsubsection{Symmetric Neumann modes}\label{symmetric_modes_Neumann_sec}
If a mode $\phi$ is symmetric about the line $x=0$, then \eqref{general_sol} implies that
\begin{equation}
\mathbf{A}^{(0)}=\mathbf{B}^{(1)}\quad\text{and}\quad\mathbf{A}^{(1)}=\mathbf{B}^{(0)}.
\end{equation}
Substituting these expressions into \eqref{S_matrix} then yields
\begin{equation}\label{symmetric_condition}
\mathbf{A}^{(0)}=\mathbf{B}^{(0)}\quad\text{and}\quad\mathbf{A}^{(1)}=\mathbf{B}^{(1)}.
\end{equation}
Since the above expression is independent of the matrix $\mathsfbi{G}$, symmetric modes are independent of the presence or absence of a barrier.

If Neumann boundary conditions are imposed on the tank walls, then combining \eqref{symmetric_condition} with \eqref{Neumann_Wall_matrix_equations} yields
\begin{equation}
    \mathsfbi{L}\mathbf{B}^{(0)}=\mathbf{B}^{(0)}.
\end{equation}
Thus $\mathbf{B}^{(0)}$ is an eigenvector of $\mathsfbi{L}$ with eigenvalue $1$. Since $\mathsfbi{L}$ is a diagonal matrix, its eigenvalues are its diagonal entries, which we recall are given by $\exp(\upi k_m W)$. The only entry that can equal unity is for $m=0$, since $k_m$ has a positive imaginary part for $m\geq 1$. The $m=0$ entry is equal to $1$ when $k_0=2l\pi/W$ for $l\in\mathbb{N}\cup\{0\}$. We denote these resonant wavenumbers as 
\begin{equation}
    \kappa_{2l}^{\mathrm{Neu}(0)}=2l\pi/W,
\end{equation} 
in which the superscript $(0)$ is included for notational consistency with \textsection\ref{multiple_barriers_sec}. The corresponding modes can be written as
\begin{equation}
    \phi_{2l}(x,z)=\cos(\kappa_{2l}^{\mathrm{Neu}(0)}x)\cosh(\kappa_{2l}^{\mathrm{Neu}(0)}(z+H)).
\end{equation}

\subsubsection{Symmetric Dirichlet modes}\label{symmetric_modes_Dirichlet_sec}
If Dirichlet boundary conditions are imposed on the tank walls, then combining \eqref{symmetric_condition} with \eqref{Dirichlet_Wall_matrix_equations} yields
\begin{equation}
    \mathsfbi{L}\mathbf{B}^{(0)}=-\mathbf{B}^{(0)}.
\end{equation}
Thus $\mathbf{B}^{(0)}$ is an eigenvector of $\mathsfbi{L}$ with eigenvalue $-1$. We find that the resonant wavenumbers are 
\begin{equation}
    \kappa_{2l-1}^{\mathrm{Dir}(0)}=(2l-1)\pi/W
    \quad\text{for}\quad 
    l\in\mathbb{N},
\end{equation} 
with the corresponding modes
\begin{equation}
    \phi_{2l-1}(x,z)=\cos(\kappa_{2l-1}^{\mathrm{Dir}(0)}x)\cosh(\kappa_{2l-1}^{\mathrm{Dir}(0)}(z+H)).
\end{equation}

\subsubsection{Antisymmetric Neumann modes}\label{antisymmetric_modes_Neumann_sec}
If a mode $\phi$ is antisymmetric about the line $x=0$, then \eqref{general_sol} implies that
\begin{equation}
\mathbf{A}^{(0)}=-\mathbf{B}^{(1)}\quad\text{and}\quad\mathbf{A}^{(1)}=-\mathbf{B}^{(0)}.
\end{equation}
Substituting these expressions into \eqref{S_matrix} then yields
\begin{equation}\label{antisymmetric_condition}
(2\mathsfbi{G}-\mathsfbi{I})\mathbf{A}^{(0)}=-\mathbf{B}^{(0)}\quad\text{and}\quad
(2\mathsfbi{G}-\mathsfbi{I})\mathbf{B}^{(1)}=-\mathbf{A}^{(1)}.
\end{equation}
In contrast with the symmetric modes, these modes depend on the submergence of the barrier. 

If Neumann boundary conditions are imposed on the tank walls, then combining \eqref{antisymmetric_condition} with \eqref{Neumann_Wall_matrix_equations} yields
\begin{equation}\label{nonlinear_eigenvalue_Neumann}
    [(2\mathsfbi{G}-\mathsfbi{I})\mathsfbi{L}+\mathsfbi{I}]\mathbf{B}^{(0)}=\mathbf{0},
\end{equation}
where $\mathbf{0}$ is the zero vector, which is a nonlinear eigenvalue problem because $\mathsfbi{G}$ and $\mathsfbi{L}$ depend on $\omega$. A solution to this problem will be sought in the form of an eigenvalue-eigenvector pair $(\omega,\mathbf{B}^{(0)})$, for a given barrier submergence $d$. The corresponding resonant wavenumbers are denoted $\kappa_{2l-1}^{\mathrm{Neu}(0)}$ for $l\in\mathbb{N}$.

\subsubsection{Antisymmetric Dirichlet modes}
\label{antisymmetric_modes_Dirichlet_sec}
If Dirichlet boundary conditions are imposed on the tank walls, then combining \eqref{antisymmetric_condition} with \eqref{Dirichlet_Wall_matrix_equations} yields
\begin{equation}\label{nonlinear_eigenvalue_Dirichlet}
    [(2\mathsfbi{G}-\mathsfbi{I})\mathsfbi{L}-\mathsfbi{I}]\mathbf{B}^{(0)}=\mathbf{0}.
\end{equation}
As before, we obtain a nonlinear eigenvalue problem to be solved for an eigenvalue-eigenvector pair $(\omega,\mathbf{B}^{(0)})$ for a given barrier submergence $d$. The corresponding resonant wavenumbers are denoted $\kappa_{2l}^{\mathrm{Dir}(0)}$ for $l\in\mathbb{N}$.

\subsection{Numerical solution of the nonlinear eigenvalue problem}\label{numerical_tool_sec}

In contrast with the nonlinear eigenvalue problems derived in \textsection\textsection\ref{symmetric_modes_Neumann_sec} and \ref{symmetric_modes_Dirichlet_sec}, those derived in \textsection\ref{antisymmetric_modes_Neumann_sec} and \ref{antisymmetric_modes_Dirichlet_sec} cannot be solved analytically because the matrix is not diagonal. Instead, we seek a numerical solution of \eqref{nonlinear_eigenvalue_Neumann} and \eqref{nonlinear_eigenvalue_Dirichlet} by employing an iterative method adapted from \cite{ooi2008analysis}. Variants of this method have previously been applied to find the complex resonances of arrays of floating bodies immersed in water \cite{WOLGAMOT2017232,chowdhury2023coupled}, vibrating ice shelves \cite{bennetts2021complex} and two-dimensional acoustic resonators \cite{wilks2023_2d_time_domain}. 
Given a nonlinear eigenvalue problem of the form
\begin{equation}
    \mathsfbi{M}(\omega)\mathbf{v}=\mathbf{0},\label{nonl_eig_problem}
\end{equation}
and initial guess $\omega_0$ of an eigenvalue $\omega$, we derive an iterative scheme by approximating \eqref{nonl_eig_problem} using a first-order Taylor series, yielding
\begin{equation}\label{nonl_eig_problem2}
    \mathsfbi{M}(\omega_0)\mathbf{v}+(\omega-\omega_0)\mathsfbi{M}^\prime(\omega_0)\mathsfbi{v}\approx 0.
\end{equation}
In \eqref{nonl_eig_problem2}, the matrix derivative is approximated by the finite difference formula
\begin{equation}
    \mathsfbi{M}^\prime(\omega)\approx\frac{\mathsfbi{M}(\omega+h)-\mathsfbi{M}(\omega)}{h},
\end{equation}
where we take $h=10^{-7}$. Equation \eqref{nonl_eig_problem2} can be rewritten as the following generalised eigenvalue problem
\begin{equation}
    \mathsfbi{M}(\omega_0)\mathbf{v}=\lambda\mathsfbi{M}^\prime(\omega_0)\mathbf{v},
\end{equation}
which we solve numerically for the $\lambda$, taken to be the smallest eigenvalue in absolute value. Then, $\omega=\omega_0-\lambda$ provides a refinement of the initial guess $\omega_0$. The method can be applied iteratively until the desired accuracy is reached.

Provided the initial guess is already close to a nonlinear eigenvalue, the method converges rapidly. We typically obtain
\begin{equation}
    \frac{\|\mathsfbi{M}(\omega)\mathbf{v}\|}{\|\mathbf{v}\|}<10^{-11}
\end{equation}
within four iterations, where $\|\cdot\|$ is the Euclidean norm. Due to its rapid convergence, the method can be applied as part of an efficient homotopy procedure. For example, if we suppose that the nonlinear eigenvalues have been found for a given value of the barrier submergence $d_0$, these eigenvalues can serve as the initial guesses for a new barrier submergence depth $d_0+\epsilon$, where $\epsilon$ is sufficiently small. Repetition of this process for $d_0+2\epsilon$, $d_0+3\epsilon$ and so on, then yields curves which show how the nonlinear eigenvalues depend on the continuous parameter $d$.

In order to initialise the homotopy method described above, we compute the antisymmetric eigenmodes of a tank without a barrier (i.e.\ with $d=0$). These eigenvalues can be derived directly from the boundary value problem \eqref{bvp_tank_appendix}, which is straightforward to solve when there is no barrier. We obtain
\begin{equation}
    \phi(x,z)=\sin(kx)\cosh(k(z+H))
\end{equation}
where 
\begin{equation}
    k=\kappa_{2l-1}^{\mathrm{Neu}(0)}=(2l-1)\pi/W
    \quad
    \text{or}
    \quad
    k=\kappa_{2l}^{\mathrm{Dir}(0)}=2l\pi/W
    \quad
    \text{for}
    \quad
    l\in\mathbb{N}
\end{equation} 
if the tank has Neumann or Dirichlet walls, respectively. Note that when $d=0$ the resonant wavenumbers of the antisymmetric modes interleave with those of the symmetric modes for both boundary condition types.

\subsection{Resonant frequencies and modes of the tank}\label{appendix_eigenvalues_sec}
\begin{figure}
    \centering
    \includegraphics[width=\textwidth]{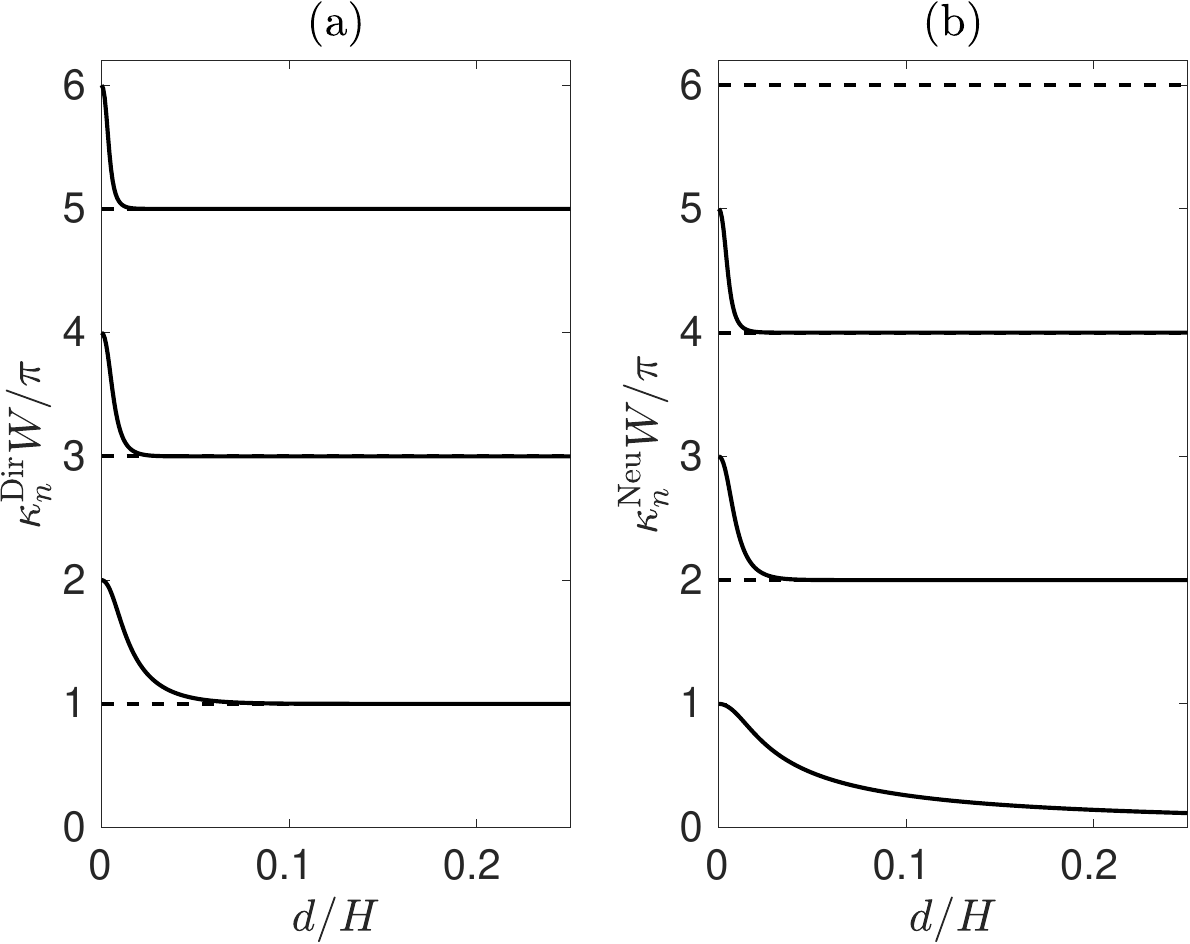}
    \caption{Evolution of the first six non-zero resonant wavenumbers in a tank containing a single barrier and with (a) Dirichlet walls and (b) Neumann walls, as a function of the barrier submergence $d$. The parameter values are $H=20$\,m and $W=2$\,m. The resonant wavenumbers of symmetric modes are plotted using dashed lines, whereas the resonant wavenumbers of antisymmetric modes are plotted using solid lines.}
    \label{fig:single_barrier_homotopy}
\end{figure}

Figure \ref{fig:single_barrier_homotopy} shows the evolution of the first six non-zero resonant wavenumbers of a Dirichlet and Neumann tank as the barrier submergence $d$ increases. In the case of the Dirichlet tank, the resonant wavenumbers of the antisymmetric modes $\kappa_{2l}^{\mathrm{Dir}(0)}$ converge towards $\kappa_{2l-1}^{\mathrm{Dir}(0)}=(2l-1)\pi/W$ as $d$ increases, for $l\in\mathbb{N}$. Similarly, in the case of the Neumann tank, the resonant wavenumbers of the antisymmetric modes $\kappa_{2l-1}^{\mathrm{Neu}(0)}$ converges towards $\kappa_{2l-2}^{\mathrm{Neu}(0)}=(2l-2)\pi/W$ as $d$ increases, for $l\in\mathbb{N}\cup\{0\}$.

The plots shown in figure \ref{fig:single_barrier_homotopy} are of the same form as those given in \cite{evans1987resonant}. We have validated our method by reproducing the results in figure 1 of that paper by using the same parameters. We obtained an excellent visual agreement after digitally overlaying our results with theirs. This comparison is not shown here for brevity.

\begin{figure}
    \centering
    \includegraphics[width=\textwidth]{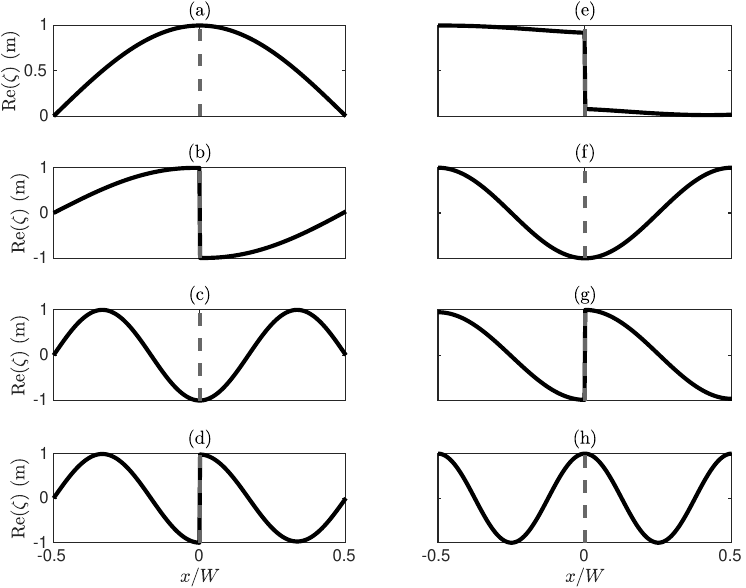}
    \caption{The real part of the free surface elevation associated with the first four nontrivial eigenmodes of the single-barrier tank with (a--d) Dirichlet walls, and (e--h) Neumann walls. The position of the barrier is marked with a vertical dashed line. The parameter values are $W=2$\,m, $H=20$\,m and $d=0.6522$\,m.}
    \label{fig:single_barrier_modes}
\end{figure}

Figure \ref{fig:single_barrier_modes} shows the real part of the free surface elevation associated with the first four non-trivial eigenmodes of a single-barrier tank, in the case of both Dirichlet and Neumann walls. Here, the complex-valued free surface elevation $\zeta$ is given by
\begin{equation}\label{free-surface-def}
    \zeta(x) = \frac{\upi\omega}{g}\phi(x,0).
\end{equation}
We observe that the symmetric modes are continuous over the barrier, whereas the antisymmetric modes are discontinuous at the location of the barrier.

\subsection{Relationship to the band structure}\label{single_band_structure_relation_sec}
The Dirichlet and Neumann eigenmodes are also solutions of the quasi-periodic Bloch problem. This problem describes wave propagation through an infinite, periodic array of vertical barriers, each with identical submergence depth $d$ and positioned at $x=nW$ for all $n\in\Z$. Bloch waves satisfy \eqref{bvp_tank_appendix} as well as the quasi-periodic boundary conditions\begin{subequations}\label{QP_BCs}
\begin{align}
\phi(W/2,z)&=\exp(\upi q W)\phi(-W/2,z)\label{QP_BC1}\\
\partial_x\phi(W/2,z)&=\exp(\upi q W)\partial_x\phi(-W/2,z),\label{QP_BC2}
\end{align}
\end{subequations}
for all $z\in[-H,0]$, where $q$ is the Bloch wavenumber.

To show that the Dirichlet and Neumann eigenmodes satisfy \eqref{QP_BCs}, we note that the Dirichlet modes trivially satisfy \eqref{QP_BC1} and the Neumann modes trivially satisfy \eqref{QP_BC2}. The remaining condition also follows as a consequence of the symmetry or antisymmetry of the mode. For example, consider the symmetric mode of the Dirichlet tank. We compute
\begin{align}
    \phi(-x,z)=\phi(x,z)\quad\implies\quad \partial_x\phi(-x,z)=-\partial_x\phi(x,z),\label{symmetric_dirichlet_bloch}
\end{align}
which follows from the chain rule. After substituting $x=W/2$ into \eqref{symmetric_dirichlet_bloch}, we observe that \eqref{QP_BC2} is satisfied with $qW=\pi$. Thus, the symmetric Dirichlet mode also describes a Bloch wave. These steps can be carried out for both wall types and for both symmetric and antisymmetric modes, as summarised in table \ref{Bloch_table}.

\begin{table}
\begin{center}
\begin{tabular}{|l|l|l|}
\hline
 & \textbf{Dirichlet walls} & \textbf{Neumann Walls} \\ \hline
\multirow{3}{*}{\textbf{\begin{tabular}[c]{@{}l@{}}Symmetric\\ modes\end{tabular}}} & $\phi(-W/2,z)=\phi(W/2,z)=0$ & $\partial_x\phi(-W/2,z)=\partial_x\phi(W/2,z)=0$ \\
 & $\partial_x\phi(-W/2,z)=-\partial_x\phi(W/2,z)$ & $\phi(-W/2,z)=\phi(W/2,z)$ \\
 & $\implies qW=\pi$ & $\implies qW=0$ \\ \hline
\multirow{3}{*}{\textbf{\begin{tabular}[c]{@{}l@{}}Antisymmetric\\ modes\end{tabular}}} & $\phi(-W/2,z)=\phi(W/2,z)=0$ & $\partial_x\phi(-W/2,z)=\partial_x\phi(W/2,z)=0$ \\
 & $\partial_x\phi(-W/2,z)=\partial_x\phi(W/2,z)$ & $\phi(-W/2,z)=-\phi(W/2,z)$ \\
 & $\implies qW=0$ & $\implies qW=\pi$ \\ \hline
\end{tabular}
\end{center}
\caption{Summary of the calculations used to show that the eigenmodes of the Dirichlet and Neumann tanks are Bloch waves through the use of symemtry.}
\label{Bloch_table}
\end{table}

\begin{figure}
    \centering
    \includegraphics[width=0.9\textwidth]{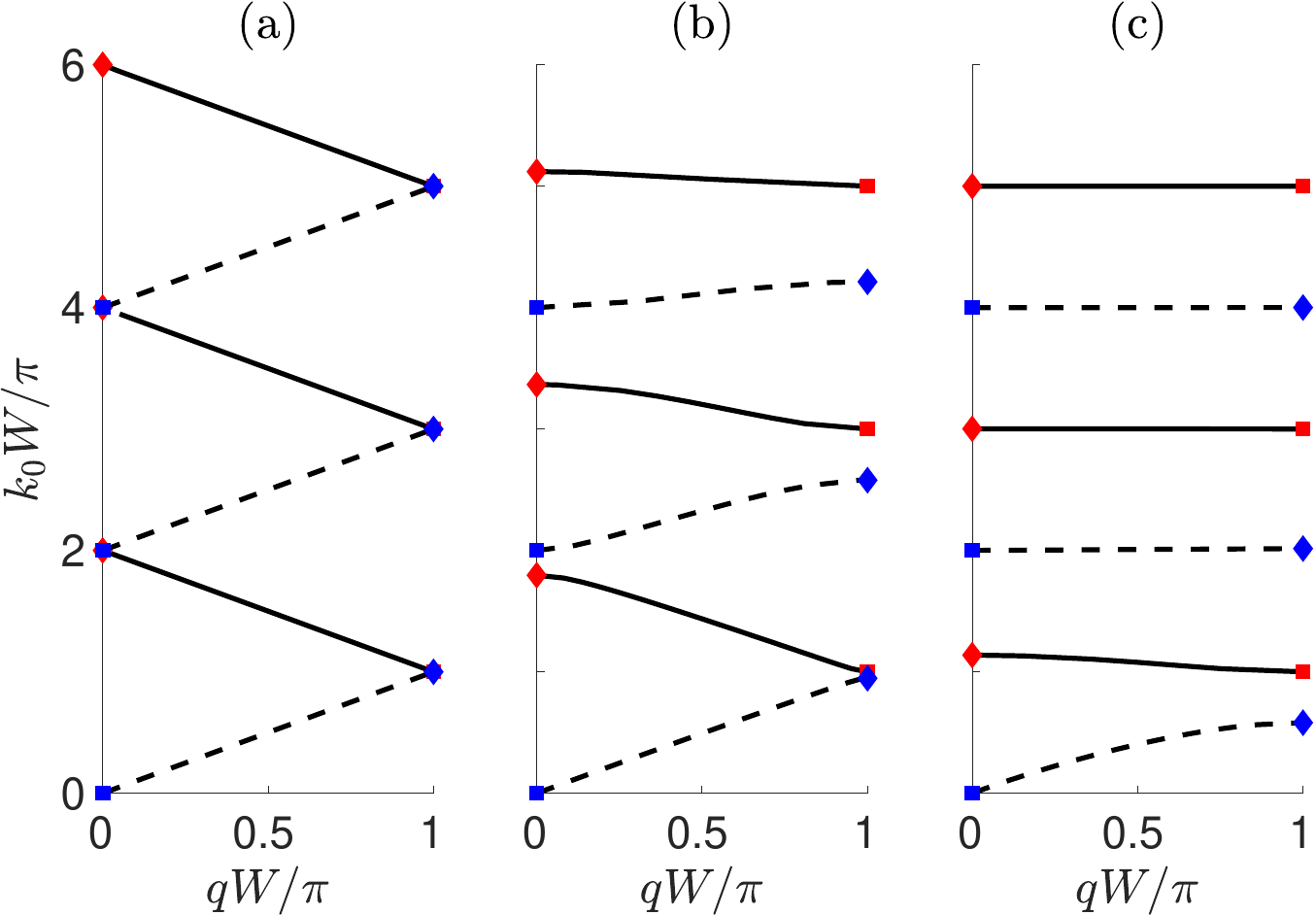}
    \caption{Band diagrams for infinite periodic arrays of vertical barriers with (a) $d=0$, (b) $d=0.1505$\,m and (c) $d=0.6522$\,m. The remaining parameters are $W=2$\,m and $H=20$\,m. The band is plotted with a solid line if it occupies the interval between two adjacent resonances of the Dirichlet tank (i.e.\ $[\kappa_{2l-1}^{\mathrm{Dir}(0)},\kappa_{2l}^{\mathrm{Dir}(0)}]$) whereas it is plotted with a dashed line if it occupies the interval between two adjacent resonances of the Neumann tank (i.e.\ $[\kappa_{2l-2}^{\mathrm{Neu}(0)},\kappa_{2l-1}^{\mathrm{Neu}(0)}]$). Periodic solutions which correspond to modes of the Dirichlet (Neumann) tank are marked in red (blue). The corresponding marker is a square (diamond) if the associated mode of the tank is symmetric (antisymmetric). Note that these markers overlap in panel (a).}
    \label{fig:passbands}
\end{figure}

Figure \ref{fig:passbands} shows the effect of $d$ on the band diagram of the periodic array. When $d=0$, the waves are unimpeded and the passband contains all wavenumbers. As $d$ increases, the infinite passband breaks at the resonant wavenumbers of the Dirichlet and Neumann tanks, which are at the edges of the band (as shown in table \ref{Bloch_table}). We observe that the passbands are confined to the intervals between two adjacent resonant wavenumbers of the corresponding Dirichlet tank, or between two adjacent resonant wavenumbers of the corresponding Neumann tank. Specifically, these intervals are $[\kappa_{2l-1}^{\mathrm{Dir}(0)},\kappa_{2l}^{\mathrm{Dir}(0)}]$ and $[\kappa_{2l-2}^{\mathrm{Neu}(0)},\kappa_{2l-1}^{\mathrm{Neu}(0)}]$, respectively. In figure \ref{fig:single_barrier_homotopy}, we observed that these resonant wavenumbers converge towards each other as $d$ increases. This effectively ``pinches'' the passbands, that is, they become increasingly thinner as $d$ increases. This pinching effect is clearly visible in figure \ref{fig:passbands}.

In \cite{wilks2022rainbow}, the authors were unable to detect the thin, high frequency passbands. This is because their method was based on frequency sweeping. That is, they discretised the frequency domain as $\omega_n=n\Delta\omega$ for $n\in\{0,\dots,N_{\omega}\}$. Then, at each $\omega_n$, they computed the corresponding Bloch wavenumber $q_n$, if such a Bloch wavenumber exists, and plotted all pairs $(q_n,\omega_n)$ in order to assemble the band diagram. This frequency sweeping method fails to detect a passband if it lies between two points of our discretisation, that is, if
\begin{subequations}\label{mesh_too_coarse}
\begin{equation}
    \omega_n<\omega(\kappa_{2l-1}^{\mathrm{Dir}(0)})<\omega(\kappa_{2l}^{\mathrm{Dir}(0)})<\omega_{n+1}
\end{equation}
in the case of a passband associated with a Dirichlet mode, or
\begin{equation}
    \omega_n<\omega(\kappa_{2l-2}^{\mathrm{Neu}(0)})<\omega(\kappa_{2l-1}^{\mathrm{Neu}(0)})<\omega_{n+1}
\end{equation}
\end{subequations}
in the case of a passband associated with a Neumann mode. In \eqref{mesh_too_coarse}, we have defined
\begin{equation}
    \omega(k)=\sqrt{gk\tanh kH}.
\end{equation}
The band diagrams in figure \ref{fig:passbands} also used this frequency sweeping method, although the Dirichlet and Neumann resonant frequencies (and some extra points between them) were added to the frequency discretisation in order to avoid the situations described in \eqref{mesh_too_coarse}. This method of augmenting the frequency discretisation yields band diagrams which are in excellent agreement with those computed using COMSOL (G. J. Chaplain, personal communication, May 17, 2023).

In \cite{wilks2022rainbow}, the cutoff frequency of the lowest frequency passband was described as being associated with the resonant frequency of a pair of identical vertical barriers in an infinite fluid domain. This resonance, which was previously studied by \cite{Newman1974, McIver1985}, is associated with a vertical fluid motion between the barriers. As we have seen, the lowest non-zero resonant wavenumber of the Neumann tank is precisely the cutoff of the lowest-frequency passband, and the corresponding resonant mode of the tank is precisely the standing Bloch wave at the cutoff. Because the pair of vertical barriers in the infinite fluid region has a somewhat similar geometry to the Neumann tank, its resonant frequency happens to be similar to the cutoff frequency. That said, the correspondence between the standing Bloch wave and the resonance between a pair of barriers is not direct. This explains why the authors of \cite{wilks2022rainbow} did not observe perfect agreement between the resonant frequency and the cutoff frequency.

\section{Multiple barriers in a tank}\label{multiple_barriers_sec}
Next, we consider the case where the tank contains $N+1$ identical vertical barriers (of uniform submergence depth $d$), whose horizontal coordinates are given by $x=nW$ for $n\in\{0,\dots,N\}$. The walls of the tank are located at $x=-W/2$ and $x=(N+\tfrac{1}{2})W$, so that the tank is of width $(N+1)W$. We consider both cases of Neumann or Dirichlet walls. The problem of finding the eigenmodes is a multiple scattering problem, which we reduce to a nonlinear eigenvalue problem of the form of \eqref{nonl_eig_problem}.

\subsection{Eigenvalue problem}
First, we partition the horizontal domain into $N+2$ intervals given by
\begin{equation}\label{intervals}
    I_n=\begin{cases}
        (-W/2,0)&n=0\\
        ((n-1)W,nW)&1\leq n\leq N\\
        (NW,(N+\tfrac{1}{2})W)&n=N+1.
    \end{cases}
\end{equation}
We then express the general solution as
\begin{equation}\label{general_sol_multiple}
    \phi(x,z)=\sum_{m=0}^\infty (A_m^{(n)}\exp(\upi k_m(x-nW))+B_m^{(n)}\exp(-\upi k_m(x-nW))\psi_m(z),
\end{equation}
for all $(x,z)\in I_n\times [-H,0]$ and for all $n\in\{0,\dots,N+1\}$.

Equation \eqref{general_sol_multiple} must satisfy the boundary conditions on the barriers. After incorporating the scattering matrix \eqref{S_matrix} and accounting for the phase shifts of the horizontal coordinates (i.e. the terms $x-nW$ appearing in the exponents in \eqref{general_sol_multiple}), we obtain
\begin{equation}\label{S_matrix2}
    \begin{bmatrix}\mathsfbi{L G}&\mathsfbi{L}(\mathsfbi{I}-\mathsfbi{G})\mathsfbi{L}\\\mathsfbi{I}-\mathsfbi{G}&\mathsfbi{GL}\end{bmatrix}
    \begin{bmatrix}
        \mathbf{A}^{(n-1)}\\\ \mathbf{B}^{(n)}
    \end{bmatrix}
    =
    \begin{bmatrix}
        \mathbf{A}^{(n)}\\\ \mathbf{B}^{(n-1)}
    \end{bmatrix},
\end{equation}
for all $n\in\{1,\dots,N+1\}$, where $\mathbf{A}^{(n)}$ and $\mathbf{B}^{(n)}$ are vectors with entries $A^{(n)}_m$ and $B^{(n)}_m$, respectively. Recall that $\mathsfbi{L}=\mathrm{diag}(e^{\upi k_m W})$ and $\mathsfbi{G}$ is obtained by solving the single-barrier scattering problem using the integral equation/Galerkin method of \cite{Porter1995a}.

The general solution \eqref{general_sol_multiple} must also satisfy the boundary conditions at the walls of the tank, $x=-W/2$ and $x=(N+\tfrac{1}{2})W$, which implies that
\begin{subequations}\label{wall_matrix_equations2}
    \begin{equation}\label{wall_matrix_eq_Neumann}
    \mathbf{B}^{(0)}=\mathsfbi{L}\mathbf{A}^{(0)}\quad\text{and}\quad
    \mathbf{A}^{(N+1)}=\mathsfbi{L}\mathbf{B}^{(N+1)},
\end{equation}
in the Neumann case, or
\begin{equation}
    \mathbf{B}^{(0)}=- \mathsfbi{L}\mathbf{A}^{(0)}\quad\text{and}\quad
    \mathbf{A}^{(N+1)}=- \mathsfbi{L}\mathbf{B}^{(N+1)},
\end{equation}
\end{subequations}
in the Dirichlet case. These conditions are analogous to \eqref{wall_matrix_equations}.

Equations \eqref{S_matrix2} and \eqref{wall_matrix_equations2} are combined into a single homogeneous system of equations of the form
\begin{equation}\label{nullspace_problem}
    \mathsfbi{M}(\omega)\begin{bmatrix}
        \mathbf{A}^{(0)}\\\vdots\\\mathbf{A}^{(N+1)}\\\mathbf{B}^{(0)}\\\vdots\\\mathbf{B}^{(N+1)}
    \end{bmatrix}=\begin{bmatrix}
        \mathbf{0}\\\vdots\\\mathbf{0}\\\mathbf{0}\\\vdots\\\mathbf{0}
    \end{bmatrix}.
\end{equation}
The matrix $\mathsfbi{M}$ is defined as
\begin{equation}\label{multiple_barriers_matrix}
    \mathsfbi{M}(\omega)=\begin{bmatrix}
    \begin{array}{ccccc|ccccc}
        \mp\mathsfbi{L}&&&&    &\mathsfbi{I}&&&&\\
        \mathsfbi{LG} &-\mathsfbi{I}&&&    &&\mathsfbi{L}(\mathsfbi{I}-\mathsfbi{G})\mathsfbi{L}&&&\\
        &\mathsfbi{LG}&\ddots&& &&&\ddots&&\\
        &&\ddots&-\mathsfbi{I}& &&&&\ddots&\\
        &&&\mathsfbi{LG}&-\mathsfbi{I} &&&&&\mathsfbi{L}(\mathsfbi{I}-\mathsfbi{G})\mathsfbi{L}\\\hline
        \mathsfbi{I}-\mathsfbi{G}&&&& &-\mathsfbi{I}&\mathsfbi{GL}&&&\\
        &\ddots&&& &&\ddots&\ddots&&\\
        &&\ddots&& &&&\ddots&\ddots&\\
        &&&\mathsfbi{I}-\mathsfbi{G}& &&&&-\mathsfbi{I}&\mathsfbi{GL}\\
        &&&&\mathsfbi{I} &&&&&\mp\mathsfbi{L}
        \end{array}
    \end{bmatrix}
\end{equation}
where taking the upper (lower) branch of the $\mp$ sign amounts to enforcing Neuman (Dirichlet) boundary conditions on the walls of the tank. Each quadrant of the matrix $\mathsfbi{M}$ (as partitioned in \eqref{multiple_barriers_matrix}) is a $(N+2)\times(N_{\mathrm{sol}}+1)$-dimensional square submatrix. Since $\mathsfbi{M}$ depends on $\omega$, equation \eqref{nullspace_problem} is a nonlinear eigenvalue problem, which we solve using the method discussed in \textsection\ref{numerical_tool_sec}.

In what follows, we let $\kappa_n^{\mathrm{Dir}(N)}$ and $\kappa_n^{\mathrm{Neu}(N)}$ denote the $n$th resonant wavenumber (in ascending order) of Dirichlet and Neumann tanks, respectively, each containing $N+1$ barriers. In particular, $\kappa_n^{\mathrm{Dir}(0)}$ and $\kappa_n^{\mathrm{Neu}(0)}$ are the resonant wavenumbers of the single-barrier tank, as described in \textsection\ref{single_barrier_tank_sec}.

\subsection{Relationship to the single-barrier problem}\label{multiple-single-sec}
As a first observation, we remark that the resonant wavenumbers of the single-barrier tank with Dirichlet (Neumann) walls are also resonant wavenumbers of the $(N+1)$-barrier tank with Dirichlet (Neumann) walls. We can see this using the following construction based on the method of images. If $\phi_{\mathrm{Neu}}$ is a symmetric mode of the single-barrier Neumann tank, then
\begin{subequations}\label{method_of_mirrors}
\begin{equation}
    \phi(x,z)\coloneqq\begin{cases}
        \phi_{\mathrm{Neu}}(x,z)&x\in[-W/2,W/2]\\
        \phi_{\mathrm{Neu}}(x-W,z)&x\in(W/2,3W/2]\\
        \quad\quad\vdots&\quad\quad\vdots\\
        \phi_{\mathrm{Neu}}(x-NW,z)&x\in((N-\tfrac{1}{2})W,(N+\tfrac{1}{2})W]
    \end{cases}
\end{equation}
is a mode of the $N+1$ barrier Neumann tank. Likewise, if $\phi_{\mathrm{Neu}}$ is an antisymmetric mode, then
\begin{equation}
    \phi(x,z)\coloneqq\begin{cases}
        \phi_{\mathrm{Neu}}(x,z)&x\in[-W/2,W/2]\\
        -\phi_{\mathrm{Neu}}(x-W,z)&x\in(W/2,3W/2]\\
        \quad\quad\vdots&\quad\quad\vdots\\
        (-1)^N\phi_{\mathrm{Neu}}(x-NW,z)&x\in((N-\tfrac{1}{2})W,(N+\tfrac{1}{2})W]
    \end{cases}
\end{equation}
\end{subequations}
is a mode of the $N+1$ barrier Neumann tank. Similar constructions show that the resonant wavenumbers of the single-barrier Dirichlet tank are also resonant frequencies of the $N+1$ barrier Dirichlet tank.

\begin{figure}
    \centering
    \includegraphics[width=\textwidth]{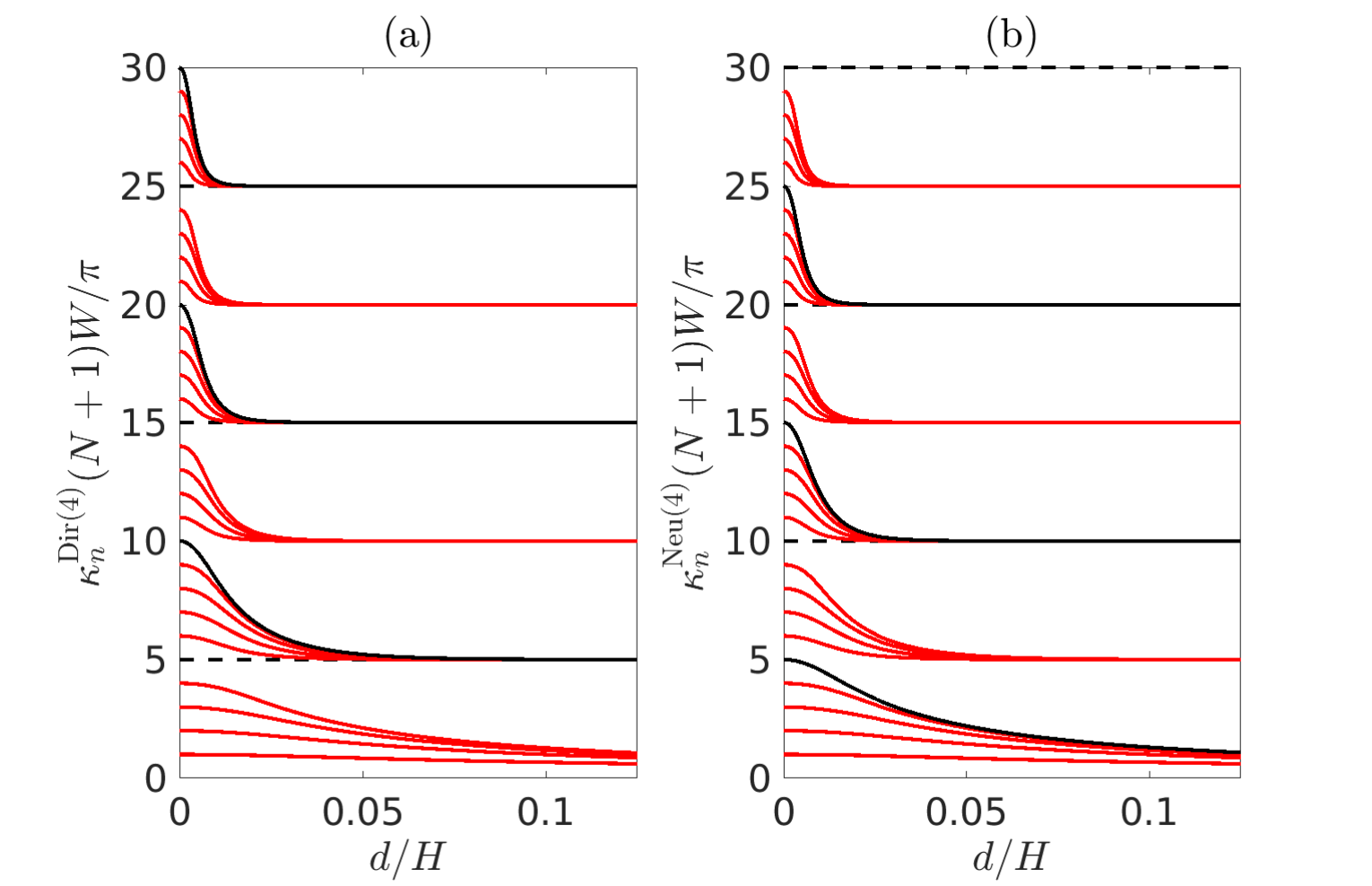}
    \caption{Evolution of the first 30 non-zero resonant wavenumbers in a tank containing $N+1=5$ barriers with (a) Dirichlet walls and (b) Neumann walls, as a function of the uniform barrier submergence $d$. The parameter values are $H=20$\,m and $W=2$\,m. Resonant wavenumbers that coincide with those of the single barrier tank with the same boundary conditions are plotted using black lines. In accordance with figure \ref{fig:single_barrier_homotopy}, these lines are solid (dashed) if the mode of the corresponding single barrier tank is symmetric (antisymmetric). Resonant wavenumbers that do not coincide with those of the single barrier tank with the same boundary conditions are plotted using red lines.}
    \label{fig:multiple_barrier_homotopy}
\end{figure}

\begin{figure}
    \centering
    \includegraphics[width=\textwidth]{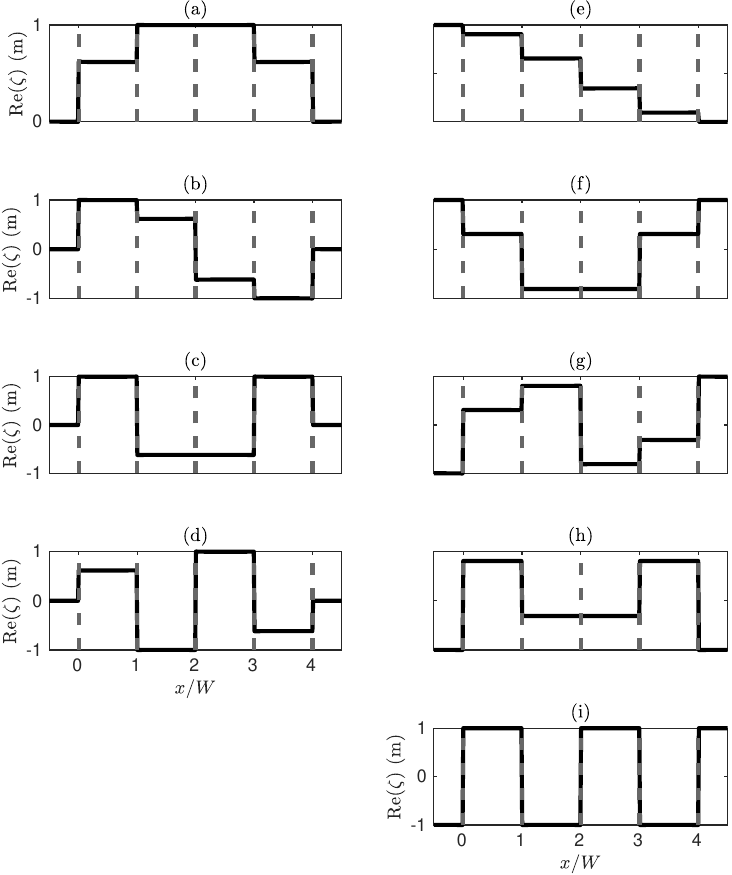}
    \caption{The real part of the free surface elevation associated with (a--d) the first four eigenmodes of the $(N+1)$-barrier Dirichlet tank, and (e--i) the first five non-trivial eigenmodes of the $(N+1)$-barrier Neumann tank. In particular, panel (i) shows the eigenmode inherited from the single-barrier Neumann tank, which is equivalent to a standing Bloch wave on the infinite array. The positions of the barriers are indicated using vertical dashed lines. The parameter values are $W=2$\,m, $H=20$\,m, $d=5$\,m and $N=4$.}
    \label{fig:multiple_barrier_modes}
\end{figure}

This effect can be observed in figure \ref{fig:multiple_barrier_homotopy}. Curves of $\kappa^{\mathrm{Dir}(0)}_n W/\pi$ and $\kappa^{\mathrm{Neu}(0)}_n W/\pi$ as functions of $d/H$ for the single barrier tank (shown in figure \ref{fig:single_barrier_homotopy}) coincide with curves of $\kappa^{\mathrm{Dir}(N)}_m W/\pi$ and $\kappa^{\mathrm{Neu}(N)}_m W/\pi$ (shown in figure \ref{fig:multiple_barrier_homotopy} for $N=4$) when $m=(N+1)n$. That is,
\begin{subequations}\label{single_to_multiple_ordering}
\begin{align}
    \kappa^{\mathrm{Dir}(0)}_n &=\kappa^{\mathrm{Dir}(N)}_{(N+1)n}\\
    \kappa^{\mathrm{Neu}(0)}_n &=\kappa^{\mathrm{Neu}(N)}_{(N+1)n}.
\end{align}
\end{subequations}
Recalling that the left-hand sides of the expressions in \eqref{single_to_multiple_ordering} delineate the edges of passbands (as described in \textsection\ref{single_band_structure_relation_sec}), these expressions are a good starting point for examining the relationship between the band structure and the modes of $(N+1)$-barrier tanks. However, the $(N+1)$-barrier tank also supports modes which are not inherited from the single-barrier tank. A collection of these are shown in figure \ref{fig:multiple_barrier_modes}, in the case where $N=4$. Our next task is to interpret these modes.

\subsection{Relationship to the band structure}
In figure \ref{fig:multiple_barrier_homotopy}(a), the resonant wavenumbers of the Dirichlet tank appear to be restricted to passbands. This can be discerned from the curves of $\kappa^{\mathrm{Dir}(N)}_m$ for $(2n-1)(N+1)\leq m\leq 2n(N+1)$, which remain between $\kappa^{\mathrm{Dir}(N)}_{(2n-1)(N+1)}$ and $\kappa^{\mathrm{Dir}(N)}_{2n(N+1)}$ as $d$ increases. Here, recall that $\kappa^{\mathrm{Dir}(N)}_{(2n-1)(N+1)}$ and $\kappa^{\mathrm{Dir}(N)}_{2n(N+1)}$ delineate the edges of a passband, which follows from the discussion in \textsection\ref{single_barrier_tank_sec}. When $\kappa^{\mathrm{Dir}(N)}_m$ is not bounded between $\kappa^{\mathrm{Dir}(N)}_{(2n-1)(N+1)}$ and $\kappa^{\mathrm{Dir}(N)}_{2n(N+1)}$ for some $n$, the resonant wavenumbers still appear to be restricted to a passband---one whose band edges are not part of the spectrum of the $(N+1)$-barrier Dirichlet tank. This is because these band edge modes are inherited from modes of the single-barrier Neumann tank, which, in turn, are not related to modes of the $(N+1)$-barrier Dirichlet tank. This can be seen by comparing panels \ref{fig:multiple_barrier_homotopy}(a) and \ref{fig:multiple_barrier_homotopy}(b), as the Neumann tank supports modes equivalent to these band edge modes. With some modification, the discussion above also describes the restriction of the resonant wavenumbers of the $(N+1)$-barrier Neumann tank to passbands.

To examine the relationship between the modes of a $(N+1)$-barrier tank and the band structure further, we focus on the lowest frequency passband and consider the effect of increasing $N$. Because $N$ varies discretely, the method described in \textsection\ref{numerical_tool_sec} is no longer effective. In particular, the resonant frequencies of a tank with $N+1$ barriers typically are not good initial guesses for the resonant frequencies of a tank with $N+2$ barriers. As an alternative, we use the following method
\begin{enumerate}
    \item For a discrete set of angular frequencies ${\omega_1<\omega_2<\dots<\omega_J}$, we compute $\det(\mathsfbi{M}(\omega_j))$ and search for intervals $(\omega_j,\omega_{j+1})$ on which both the real and imaginary parts of the determinant changes sign, i.e.
    \begin{subequations}
    \begin{align}
       \Re(\det(\mathsfbi{M}(\omega_j)))\Re(\det(\mathsfbi{M}(\omega_{j+1})))&\leq 0&\text{and}\\
       \Im(\det(\mathsfbi{M}(\omega_j)))\Im(\det(\mathsfbi{M}(\omega_{j+1})))&\leq 0.
    \end{align}
    \end{subequations}

    \item We refine the interval $[\omega_j,\omega_{j+1}]$ by applying the MATLAB built-in function \textit{fzero} to $\Re(\det(\mathsfbi{M}))$. This function, which is an implementation of the Brent-Dekker root-finding algorithm, finds a root $\omega\in[\omega_j,\omega_{j+1}]$ of $\Re(\det(\mathsfbi{M}))$, which is accurate to machine precision. Subsequently, we verify that $\Im(\det(\mathsfbi{M}(\omega)))=0$ to machine precision as well.
\end{enumerate}

\begin{figure}
    \centering
    \includegraphics[width=\textwidth]{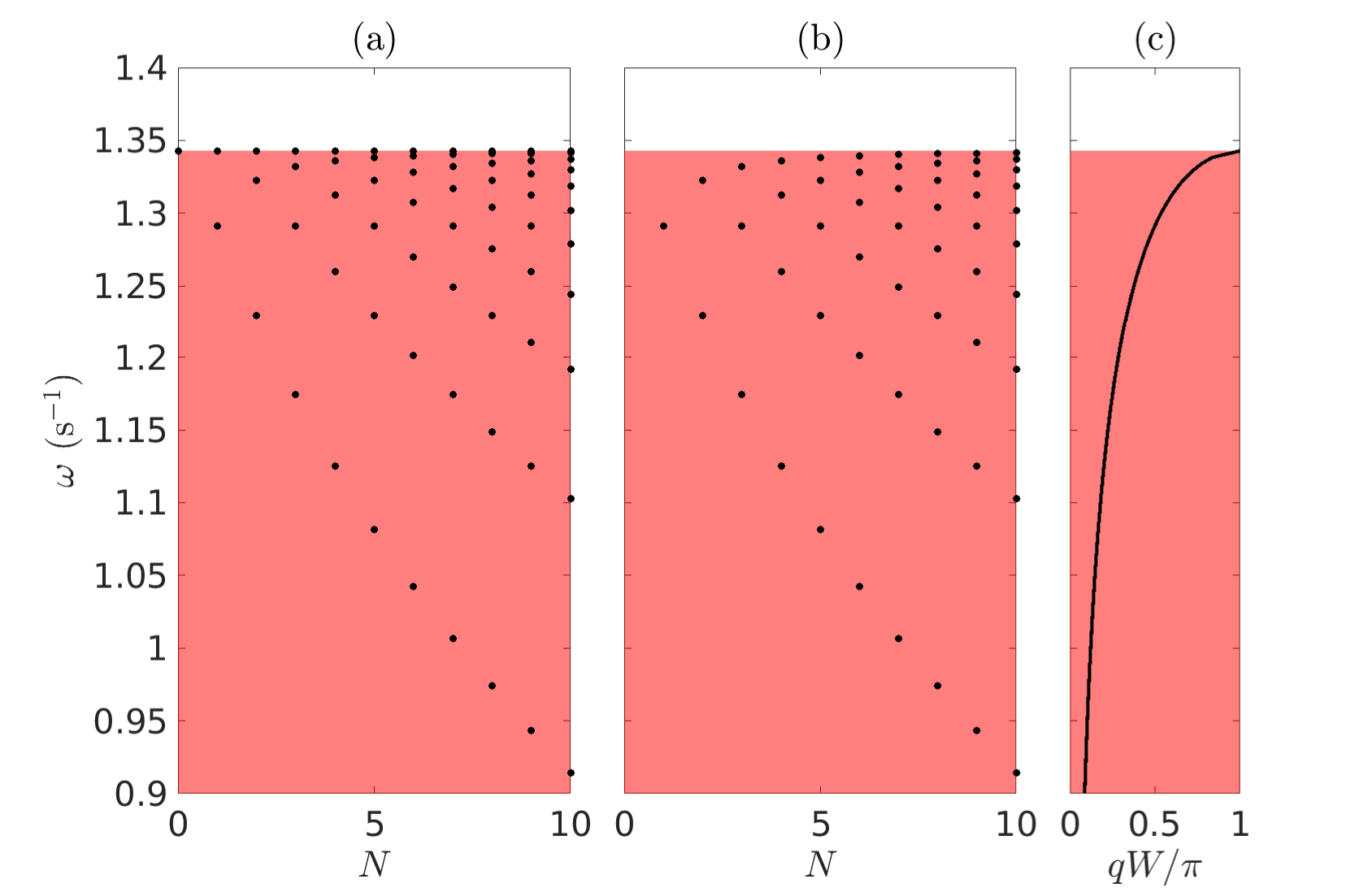}
    \caption{(Panels a--b) Resonant frequencies of a $(N+1)$-barrier tank with (a) Neumann walls, and (b) Dirichlet walls. The parameter values are $H=20$\,m, $d=5$\,m and $W=2$\,m. The passband of the corresponding infinite array is underlaid in pink. The band diagram of the corresponding infinite array is shown in panel (c) in the $q$-$\omega$ plane.}
    \label{fig:Effect_of_N}
\end{figure}

In figure \ref{fig:Effect_of_N}(a), we observe that in the case of the Neumann tank, the cutoff frequency of the lowest frequency passband is always a part of the spectrum, which follows from the method of images construction \eqref{method_of_mirrors}. Moreover, we find $N$ resonant frequencies below this cutoff frequency. In the case of the Dirichlet tank (figure \ref{fig:Effect_of_N}b), the passband only supports $N$ resonant frequencies. The cutoff frequency is not a resonant frequency of the $(N+1)$-barrier Dirichlet tank because it is not a resonant frequency of the single-barrier Dirichlet tank. In the case of both Dirichlet and Neumann tanks, the number of resonant frequencies in the passband increases as $N$ increases, which eventually form a discrete approximation of the passband. In other words, the resonant frequencies become increasingly densely packed within the passband as the number of barriers spanning the tank increases.

\begin{figure}
    \centering
    \includegraphics[width=\textwidth]{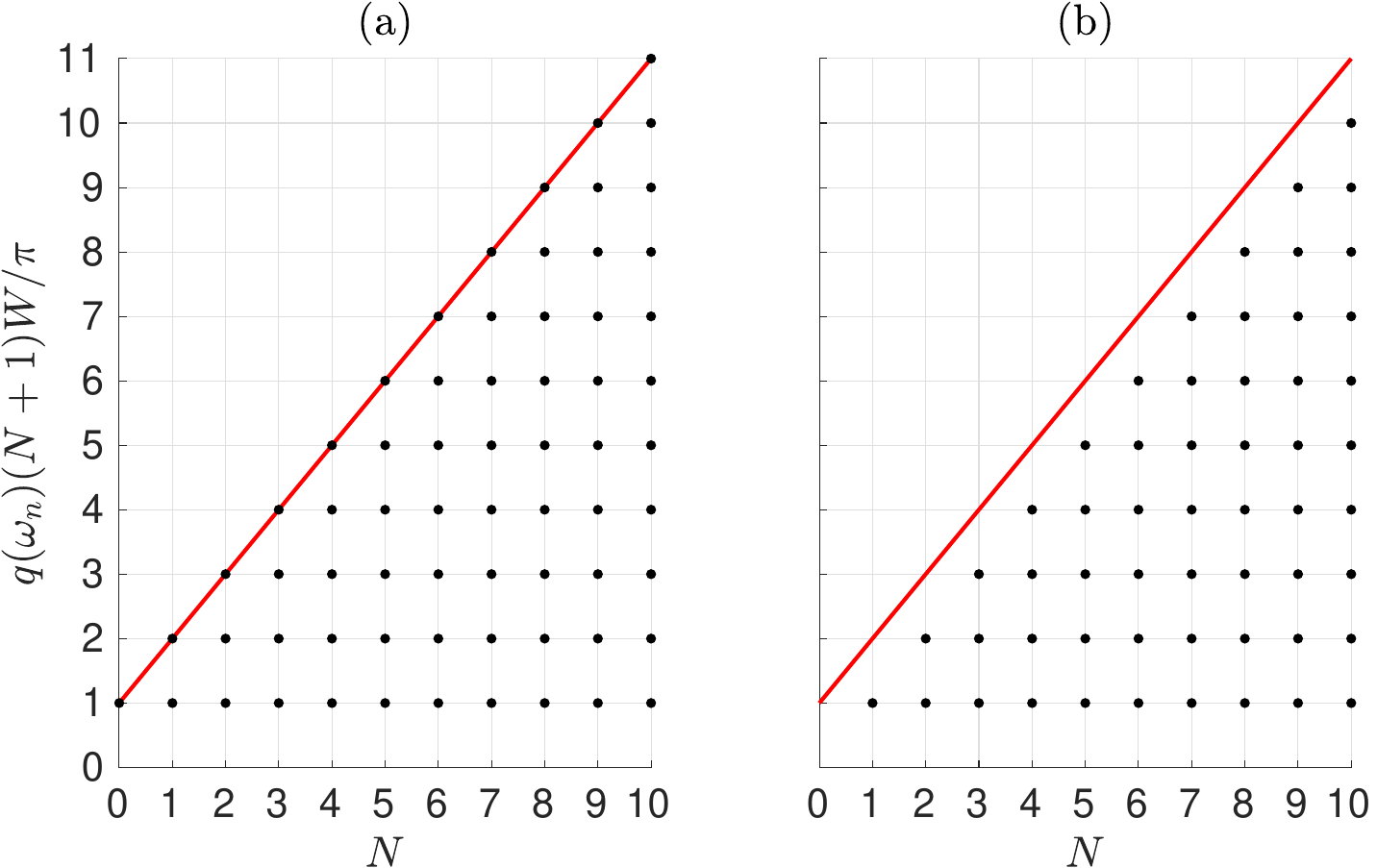}
    \caption{Plots showing twice the number of Bloch wavelengths spanning the tank at the resonant frequencies (black dots), for the tank with (a) Neumann walls, and (b) Dirichlet walls. The line $qW=\pi$, which describes the cutoff of the lowest frequency passband, is shown with a red line.}
    \label{fig:Bloch_wavenumber_integer}
\end{figure}

We claim that the modes of the $(N+1)$-barrier tanks can be thought of as a superposition of left- and right-travelling Bloch waves, which reflect off the walls of the tank and undergo constructive interference. To demonstrate this, figure \ref{fig:Bloch_wavenumber_integer} shows plots of $q(\omega_n)(N+1)W/\pi$ as $N$ increases. In particular, for each resonant frequency $\omega_n$, we compute the Bloch wavenumber $q(\omega_n)$ of the corresponding infinite array. The quantity $q(\omega_n)(N+1)W/\pi$ can be interpreted as twice the length of the tank divided by the Bloch wavelength $2\pi/q(\omega_n)$. Since figure \ref{fig:Bloch_wavenumber_integer} shows that these quantities coincide almost perfectly with the integer grid, this supports our claim that the resonant modes consist of counter-propagating Bloch waves.

\section{Quasimodes of an array of barriers}\label{quasimodes_sec}
We now consider the problem of water wave scattering by $N+1$ identical barriers in a fluid of infinite horizontal extent. This problem can be obtained by modifying \eqref{intervals} so that $I_0$ and $I_{N+1}$ become semi-infinite regions, namely
\begin{align}
    I_0=(-\infty,0)\quad\text{and}\quad I_{N+1}=(NW,\infty).
\end{align}
In other words, the walls of the tank have been removed. The Dirichlet or Neumann boundary conditions on the walls are replaced by the Sommerfeld radiation condition
\begin{equation}
     \left(\frac{\partial}{\partial x} \mp \upi k_0\right)(\phi-\phi_{\mathrm{Inc}})\to 0\quad\mbox{as\ }k_0x\to\pm\infty\label{sommerfeld}
\end{equation}
where $\phi_{\mathrm{Inc}}$ is a prescribed incident wave. Although the scattering problem can be formulated by writing a matrix equation similar to \eqref{nullspace_problem} with a non-zero right-hand-side vector due to the forcing $\phi_{\mathrm{Inc}}$, we instead solve it using the scattering matrix method developed in \cite{wilks2022rainbow}. This scattering matrix method, which is based on earlier work by \cite{Ko1988,Bennetts2009}, reduces the $(N+1)\times(N_{\mathrm{sol}}+1)$ dimensional system into $N+1$ problems, each of dimension $N_{\mathrm{sol}}+1$.

We consider the relationship between the resonant modes of a $(N+1)$-barrier tank and the quasimodes of $N+1$ barriers in a domain of infinite horizontal extent. These quasimodes are purely outgoing solutions of the scattering problem which exist for a discrete set of complex frequencies. With reference to \eqref{general_sol_multiple}, the condition that the complex resonance is purely outgoing can be stated as
\begin{equation}\label{complex_resonance_outgoing_condition}
    \mathbf{A}^{(0)}=\mathbf{B}^{(N+1)}=\mathbf{0}.
\end{equation}
Motivated by the similarity between these conditions and \eqref{wall_matrix_eq_Neumann} (which arose from the Neumann boundary condition on the tank walls), we propose a homotopy procedure to illustrate the connection between the modes of a $N+1$ barrier Neumann tank and an array of $N+1$ barriers in open water. The Dirichlet tank is not considered in this section because it supports one fewer mode in the frequency interval of interest, but it would be straightforward to adapt our method to this case. We write
\begin{equation}\label{homotopy_condition}
        (1-\hbar)\mathbf{B}^{(0)}=\mathsfbi{L}\mathbf{A}^{(0)}\quad\text{and}\quad
    (1-\hbar)\mathbf{A}^{(N+1)}=\mathsfbi{L}\mathbf{B}^{(N+1)},
\end{equation}
where $\hbar\in[0,1]$ is an introduced homotopy parameter. This parameter has the effect that as $\hbar\to 0$, we recover \eqref{wall_matrix_eq_Neumann}, which is equivalent to the Neumann boundary condition on the walls of the tank. Moreover, as $\hbar\to 1$, we recover \eqref{complex_resonance_outgoing_condition}, which is equivalent to the requirement that quasimodes be purely outgoing in the far field. Intermediate values of $\hbar\in(0,1)$ should therefore interpolate between these two cases. Such values could be interpreted as defining tanks with `porous' boundaries that allow some but not all of the energy to leak outwards. That said, we note that this porosity is artificial, i.e.\ the parameter $\hbar$ does not describe the physical property of porosity (in contrast with the models studied by \cite{sahoo2000trapping}).

\begin{figure}
    \centering
    \includegraphics[width=\textwidth]{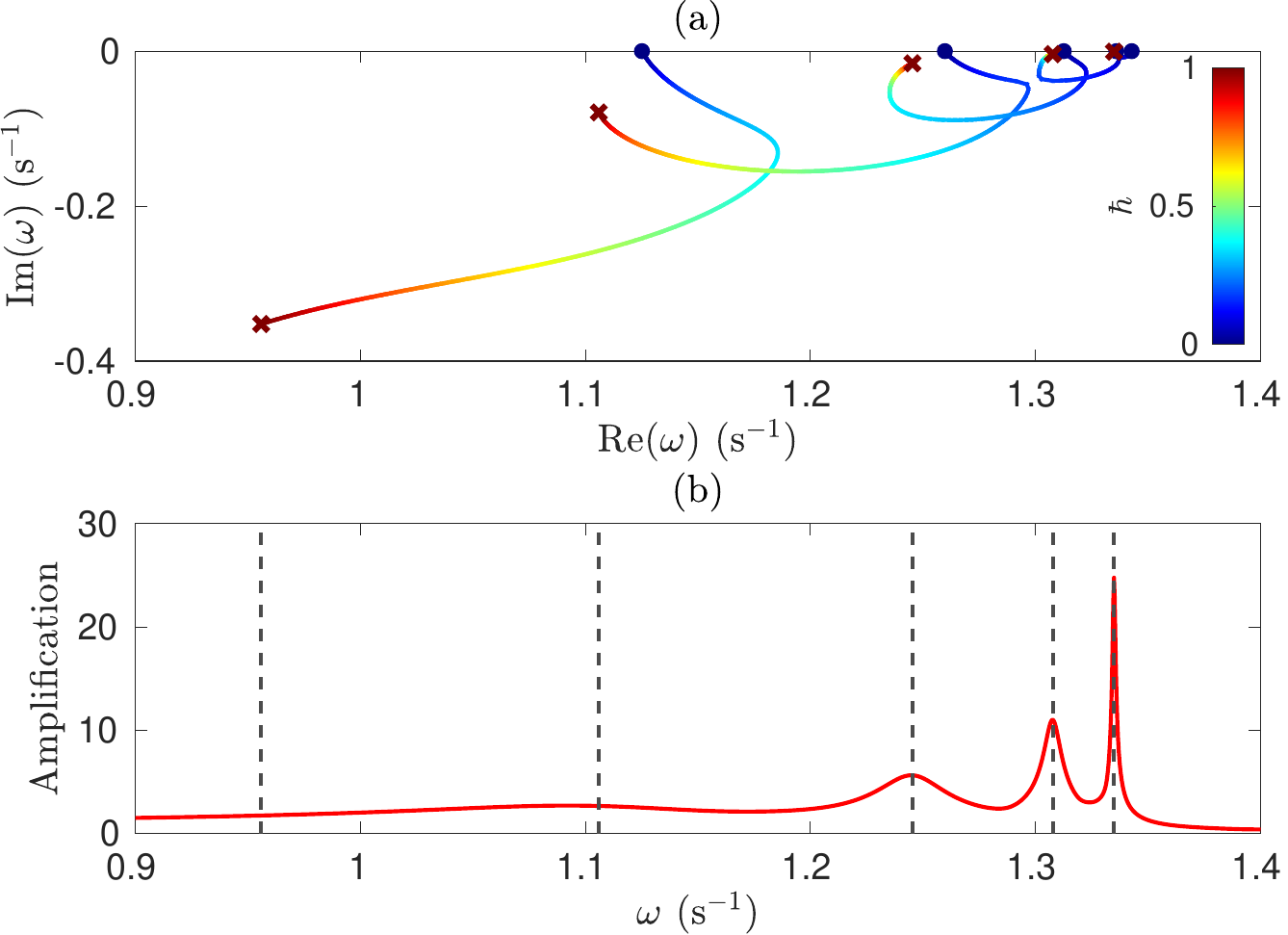}
    \caption{(a) The evolution of the first five non-trivial resonant frequencies as the homotopy parameter $\hbar$ evolves from $0$ to $1$. The value of $\hbar$ along the curves is indicated using the colour scale. The terminal points of the homotopy curves, corresponding to the resonant frequencies of the Neumann tank (for $\hbar=0$) and the complex resonant frequencies of the open problem (for $\hbar=1$), are marked as blue bullets and red crosses, respectively. (b) The amplification factor defined in \eqref{amplification} as a function of frequency. The real parts of the complex resonant frequencies are marked with dashed vertical lines. In both panels (a) and (b), the parameters are $N=4$, $H=20$\,m, $d=5$\,m and $W=2$\,m.}
    \label{fig:resonance_forcing}
\end{figure}

To compute the homotopy curves, we first incorporate \eqref{homotopy_condition} into \eqref{multiple_barriers_matrix}. The matrix $\mathsfbi{M}$ becomes
\begin{equation}\label{multiple_barriers_matrix_homotopy}
    \mathsfbi{M}(\omega;\hbar)=\begin{bmatrix}
    \begin{array}{ccccc|ccccc}
        -\mathsfbi{L}&&&&    &(1-\hbar)\mathsfbi{I}&&&&\\
        \mathsfbi{LG} &-\mathsfbi{I}&&&    &&\mathsfbi{L}(\mathsfbi{I}-\mathsfbi{G})\mathsfbi{L}&&&\\
        &\mathsfbi{LG}&\ddots&& &&&\ddots&&\\
        &&\ddots&-\mathsfbi{I}& &&&&\ddots&\\
        &&&\mathsfbi{LG}&-\mathsfbi{I} &&&&&\mathsfbi{L}(\mathsfbi{I}-\mathsfbi{G})\mathsfbi{L}\\\hline
        \mathsfbi{I}-\mathsfbi{G}&&&& &-\mathsfbi{I}&\mathsfbi{GL}&&&\\
        &\ddots&&& &&\ddots&\ddots&&\\
        &&\ddots&& &&&\ddots&\ddots&\\
        &&&\mathsfbi{I}-\mathsfbi{G}& &&&&-\mathsfbi{I}&\mathsfbi{GL}\\
        &&&&(1-\hbar)\mathsfbi{I} &&&&&-\mathsfbi{L}
        \end{array}
    \end{bmatrix},
\end{equation}
which satisfies the following nonlinear eigenvalue problem
\begin{equation}\label{nonlinear_eigenvalue2}
    \mathsfbi{M}(\omega;\hbar)\begin{bmatrix}
        \mathbf{A}^{(0)}\\\vdots\\\mathbf{A}^{(N+1)}\\\mathbf{B}^{(0)}\\\vdots\\\mathbf{B}^{(N+1)}
    \end{bmatrix}=\begin{bmatrix}
        \mathbf{0}\\\vdots\\\mathbf{0}\\\mathbf{0}\\\vdots\\\mathbf{0}
    \end{bmatrix}.
\end{equation}
Then, we let ${0=\hbar_0<\hbar_1<\dots<\hbar_M=1}$ be a discretisation of the interval $[0,1]$. We assume that the nonlinear eigenvalues are known for the Neumann tank case ($\hbar=0$), which we find by using the method described in \textsection\ref{numerical_tool_sec}. Then, the nonlinear eigenvalues of the case $\hbar=\hbar_1$ are found using the numerical method of \cite{WOLGAMOT2017232} described in \textsection\ref{numerical_tool_sec}, in which the solution for $\hbar=0$ provides the initial guess. This procedure is repeated, each time using the result of the previous iteration as the initial guess of the current iteration, until the $\hbar=1$ case is obtained. The method yields curves of the resonant frequencies as a function of $\hbar$ in the complex frequency plane, as shown in figure \ref{fig:resonance_forcing}(a).

Figure \ref{fig:resonance_forcing} illustrates the relationship between the complex resonant frequencies and the response of the array to a right-travelling incident plane wave. The free surface elevation of this incident wave is taken to be
\begin{equation}\label{incident_wave}
    \zeta_{\mathrm{Inc}}(x) = A_{\mathrm{Inc}}e^{\upi kx},
\end{equation}
where $A_{\mathrm{Inc}}$ is the prescribed amplitude of the wave. Letting $\zeta$ be the free surface elevation of the solution to this non-homogeneous scattering problem, the response of the array is quantified as
\begin{equation}\label{amplification}
    \mathrm{Amplification\ }=\frac{1}{NW}\int_0^{NW} \frac{|\zeta(x)|}{A_{\mathrm{Inc}}}\upd x,
\end{equation}
i.e., it is the ratio of the free surface amplitude to the incident amplitude, averaged spatially inside the array. In figure \ref{fig:resonance_forcing}(b), we observe that the peaks in this quantity coincide with the real parts of the complex resonant frequencies. Moreover, the height and sharpness of the peaks are inversely proportional to the imaginary part of the corresponding complex resonance---this observation can be formalised using the quality factor  \cite{pagneux2013trapped}.

\begin{figure}
    \centering
    \includegraphics[width=\textwidth]{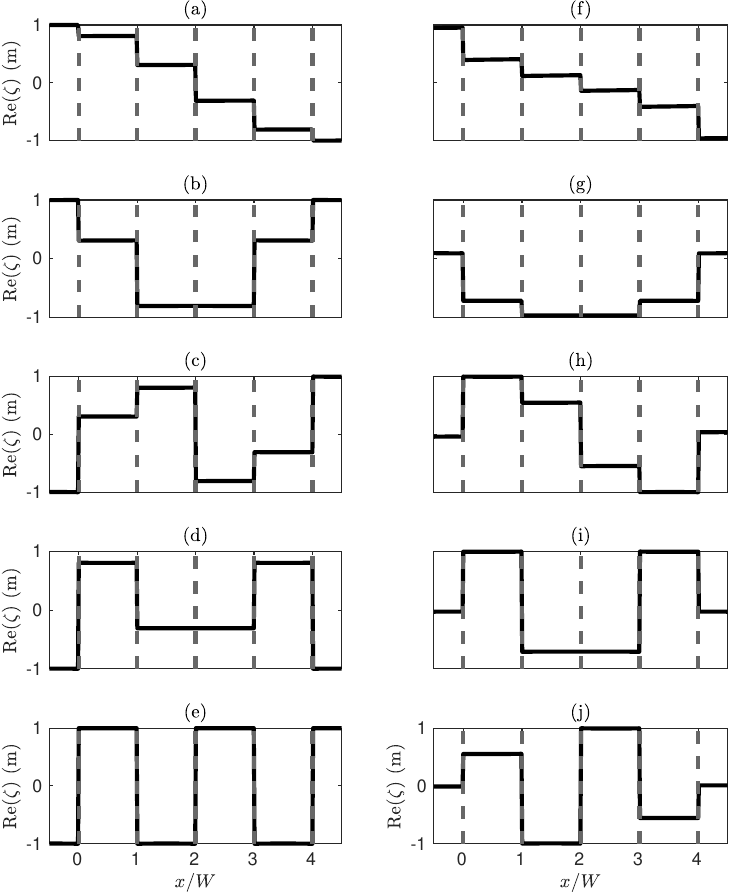}
    \caption{(f--j) The real part of the free surface elevation associated with the quasimodes of $(N+1)$ barriers in a fluid of infinite horizontal extent, which were obtained using the homotopy procedure. (a--e) To aid comparison, the corresponding modes of the Neumann tank are reproduced from figure \ref{fig:multiple_barrier_modes}(e--i), such that each row represents the terminal points of a homotopy curve in figure \ref{fig:resonance_forcing}(a). The parameter values are $W=2$\,m, $H=20$\,m and $d=5$\,m.}
    \label{fig:quasimodes}
\end{figure}

Figure \ref{fig:quasimodes} compares the modes of the $(N+1)$-barrier Neumann tank with the quasimodes of an array of $N+1$ barriers in a fluid of infinite horizontal extent. Typically, we find that there is a qualitative agreement between the modes and the corresponding quasimodes. Thus, the homotopy described by \eqref{nonl_eig_problem2} and pictured in figure \ref{fig:resonance_forcing}(a) appears to preserve the mode shapes. That said, there are differences between the tank eigenmodes and the corresponding quasimodes in figure \ref{fig:quasimodes}, which reflect the different underlying problems. These differences are genuine, because the phase and amplitudes of the modes in figure \ref{fig:quasimodes} have been chosen to maximise the qualitative agreement.

\begin{figure}
    \centering
    \includegraphics[width=\textwidth]{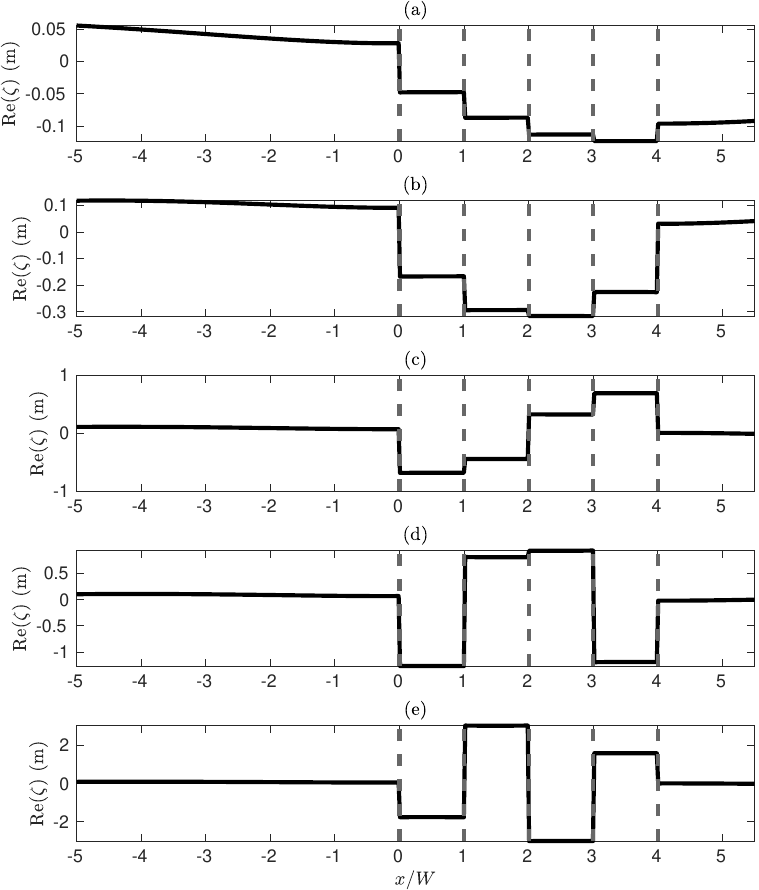}
    \caption{The real part of the free surface elevation which arises when the array of $(N+1)$-barriers immersed in a fluid of infinite horizontal extent is forced by a plane wave of the form of \eqref{incident_wave}. The frequencies of the incident waves are (a) 0.9559\,s$^{-1}$, (b) 1.1060\,s$^{-1}$, (c) 1.2453\,s$^{-1}$, (d) 1.3077\,s$^{-1}$ and (e) 1.3349\,s$^{-1}$, which are the real parts of the complex resonant frequencies obtained using the homotopy procedure. The remaining parameter values are $A_{\mathrm{Inc}}=0.1$\,m, $W=2$\,m, $H=20$\,m and $d=5$\,m.}
    \label{fig:free_surface_forcing}
\end{figure}

Figure \ref{fig:free_surface_forcing} shows the response of an array of $N+1$ barriers in a fluid of infinite horizontal extent, when the frequency of the incident wave is equal to the real part of the complex resonant frequencies. Typically, qualitative features of the mode shapes of the complex resonances, and by extension the mode shapes of the corresponding Neumann tank, can be discerned in the forcing response patterns. In other words, the nature of the response of the array to forcing is inherited from the modes of the $(N+1)$-barrier tank, with the homotopy and the quasimodes acting as the bridge between them.

\section{Conclusion}\label{conclusion_sec}
In this paper, we have explored the resonant modes of a rectangular tank containing $N+1$ vertical barriers, and the connections between these modes, the Bloch waves supported by a periodic array of barriers, and the quasimodes of a finite uniform array. 
In \textsection\ref{single_barrier_tank_sec}, we showed that the modes of the single barrier tank are equivalent to standing Bloch waves of the periodic array. Thus, their resonant frequencies delineate the edges of the passbands. As the barrier submergence increases, pairs of these resonant frequencies move closer together, which pinches the passbands---an effect which becomes stronger at higher frequencies. In \textsection\ref{multiple_barriers_sec}, we extended the model for the case $N>0$. 
We showed that the $N$ lowest resonant frequencies form a discrete approximation of the passband, and argued that the corresponding modes consist of a superposition of forwards and backwards propagating Bloch waves. In \textsection\ref{quasimodes_sec}, we introduced a homotopy procedure to relate the resonances of the $N+1$ barrier tank to the quasimodes of an array of $N+1$ barriers, and, therefore, to the resonant response to plane wave forcing.

Future work could extend this work to study of graded arrays of vertical barriers. In particular, one could explore how the resonant phenomenon of rainbow reflection is inherited from the related system in which the graded array is contained in a tank. 
Additionally, one could also seek to apply the homotopy procedure described here to line arrays of cylinders. Thus, one could formally connect the trapped modes around a line array of cylinders spanning a waveguide (e.g.\ \cite{utsunomiya_taylor_1999}) with the quasimodes of the same array in open water (e.g.\ \cite{taylor2007modelling}). This problem is more challenging than the one we have considered, because, unlike the rectangular tank, the waveguide also supports a continuous spectrum of scattering solutions, which one would need to separate from the discrete spectrum. 

\section*{Acknowledgements}
The authors would like to thank Dr Gregory J.\ Chaplain for several helpful discussions during the preparation of this manuscript.

BW, FM and LB would like to thank the Isaac Newton Institute for Mathematical Sciences at Cambridge University (UK) for their support during the programme Mathematical theory and applications of multiple wave scattering, where work on this paper was initiated.  This work was supported by EPSRC grant no EP/R014604/1. BW and FM were supported by a Simons Foundation visiting fellowship to attend the programme.

BW acknowledges financial support from a University of Otago postgraduate publishing bursary.

LB is funded by the Australian Research Council (DP200102828, FT190100404, LP180101109).

\bibliographystyle{elsarticle-num} 
\bibliography{bibfile}

\begin{thebibliography}{10}
\expandafter\ifx\csname url\endcsname\relax
  \def\url#1{\texttt{#1}}\fi
\expandafter\ifx\csname urlprefix\endcsname\relax\def\urlprefix{URL }\fi
\expandafter\ifx\csname href\endcsname\relax
  \def\href#1#2{#2} \def\path#1{#1}\fi

\bibitem{Hu2004}
X.~Hu, Y.~Shen, X.~Liu, R.~Fu, J.~Zi, {Superlensing effect in liquid surface waves}, Physical Review E - Statistical, Nonlinear, and Soft Matter Physics 69~(3 1) (2004) 030201.
\newblock \href {https://doi.org/10.1103/PhysRevE.69.030201} {\path{doi:10.1103/PhysRevE.69.030201}}.

\bibitem{Hu2011}
X.~Hu, C.~T. Chan, K.~M. Ho, J.~Zi, {Negative effective gravity in water waves by periodic resonator arrays}, Physical Review Letters 106~(17) (2011) 174501.
\newblock \href {https://doi.org/10.1103/PhysRevLett.106.174501} {\path{doi:10.1103/PhysRevLett.106.174501}}.

\bibitem{Dupont2017}
G.~Dupont, F.~Remy, O.~Kimmoun, B.~Molin, S.~Guenneau, S.~Enoch, {Type of dike using C-shaped vertical cylinders}, Physical Review B 96~(18) (2017) 180302.
\newblock \href {https://doi.org/10.1103/PhysRevB.96.180302} {\path{doi:10.1103/PhysRevB.96.180302}}.

\bibitem{Bennetts2018}
L.~G. Bennetts, M.~Peter, R.~V. Craster, {Graded resonator arrays for spatial frequency separation and amplification of water waves}, Journal of Fluid Mechanics 854 (2018) R4.
\newblock \href {https://doi.org/10.1017/jfm.2018.648} {\path{doi:10.1017/jfm.2018.648}}.

\bibitem{wilks2022rainbow}
B.~Wilks, F.~Montiel, S.~Wakes, Rainbow reflection and broadband energy absorption of water waves by graded arrays of vertical barriers, Journal of Fluid Mechanics 941 (2022) A26.
\newblock \href {https://doi.org/10.1017/jfm.2022.302} {\path{doi:10.1017/jfm.2022.302}}.

\bibitem{huang2023Water}
J.~Huang, R.~Porter, Water wave propagation through arrays of closely spaced surface-piercing vertical barriers, Journal of Fluid Mechanics 960 (2023) A20.
\newblock \href {https://doi.org/10.1017/jfm.2023.207} {\path{doi:10.1017/jfm.2023.207}}.

\bibitem{wilks_montiel_wakes_2023}
B.~Wilks, F.~Montiel, S.~Wakes, A mechanistic evaluation of the local bloch wave approximation in graded arrays of vertical barriers, Journal of Fluid Mechanics 967 (2023) A20.
\newblock \href {https://doi.org/10.1017/jfm.2023.466} {\path{doi:10.1017/jfm.2023.466}}.

\bibitem{hu2013experimental}
X.~Hu, J.~Yang, J.~Zi, C.~T. Chan, K.-M. Ho, Experimental observation of negative effective gravity in water waves, Scientific reports 3~(1) (2013) 1916.
\newblock \href {https://doi.org/10.1038/srep01916} {\path{doi:10.1038/srep01916}}.

\bibitem{Archer2020}
A.~J. Archer, H.~A. Wolgamot, J.~Orszaghova, L.~G. Bennetts, M.~A. Peter, R.~V. Craster, {Experimental realization of broadband control of water-wave-energy amplification in chirped arrays}, Phys. Rev. Fluids 5~(6) (2020) 62801.
\newblock \href {https://doi.org/10.1103/PhysRevFluids.5.062801} {\path{doi:10.1103/PhysRevFluids.5.062801}}.

\bibitem{euve2021control}
L.-P. Euv{\'e}, N.~Piesniewska, A.~Maurel, K.~Pham, P.~Petitjeans, V.~Pagneux, Control of the swell by an array of helmholtz resonators, Crystals 11~(5) (2021) 520.
\newblock \href {https://doi.org/10.3390/cryst11050520} {\path{doi:10.3390/cryst11050520}}.

\bibitem{euve2023negative}
L.-P. Euv{\'e}, K.~Pham, A.~Maurel, Negative refraction of water waves by hyperbolic metamaterials, Journal of Fluid Mechanics 961 (2023) A16.
\newblock \href {https://doi.org/10.1017/jfm.2023.220} {\path{doi:10.1017/jfm.2023.220}}.

\bibitem{zheng2020wave}
S.~Zheng, R.~Porter, D.~Greaves, Wave scattering by an array of metamaterial cylinders, Journal of Fluid Mechanics 903 (2020) A50.
\newblock \href {https://doi.org/10.1017/jfm.2020.660} {\path{doi:10.1017/jfm.2020.660}}.

\bibitem{PORTER2021102673}
R.~Porter, Plate arrays as a perfectly-transmitting negative-refraction metamaterial, Wave Motion 100 (2021) 102673.
\newblock \href {https://doi.org/10.1016/j.wavemoti.2020.102673} {\path{doi:10.1016/j.wavemoti.2020.102673}}.

\bibitem{zheng2022}
S.~Zheng, R.~Porter, H.~Liang, D.~Greaves, Water wave interaction with an annular metamaterial cylinder, in: A.~Iafrati, G.~Colicchio (Eds.), Proceedings of the 37th {International Workshop on Water Waves and Floating Bodies}, Institute of Marine Engineering, 2022, pp. 194--197.

\bibitem{huang2023surface}
J.~Huang, R.~Porter, S.~Zheng, A surface-piercing truncated cylindrical meta-structure operating as a wave energy converter, Physics of Fluids 35~(9) (2023).
\newblock \href {https://doi.org/10.1063/5.0165068} {\path{doi:10.1063/5.0165068}}.

\bibitem{zheng2024wave}
S.~Zheng, H.~Liang, D.~Greaves, Wave scattering and radiation by a surface-piercing vertical truncated metamaterial cylinder, Journal of Fluid Mechanics 983 (2024) A7.
\newblock \href {https://doi.org/10.1017/jfm.2024.147} {\path{doi:10.1017/jfm.2024.147}}.

\bibitem{Ursell1947}
F.~Ursell, {The effect of a fixed vertical barrier on surface waves in deep water}, Mathematical Proceedings of the Cambridge Philosophical Society 43~(3) (1947) 374--382.
\newblock \href {https://doi.org/10.1017/S0305004100023604} {\path{doi:10.1017/S0305004100023604}}.

\bibitem{Evans1972}
D.~V. Evans, C.~A.~N. Morris, {Complementary Approximations to the Solution of a Problem in Water Waves}, IMA Journal of Applied Mathematics 10~(1) (1972) 1--9.
\newblock \href {https://doi.org/10.1093/imamat/10.1.1} {\path{doi:10.1093/imamat/10.1.1}}.

\bibitem{Newman1974}
J.~N. Newman, Interaction of water waves with two closely spaced vertical obstacles, Journal of Fluid Mechanics 66~(1) (1974) 97--106.
\newblock \href {https://doi.org/10.1017/S0022112074000085} {\path{doi:10.1017/S0022112074000085}}.

\bibitem{Porter1995a}
R.~Porter, D.~V. Evans, {Complementary approximations to wave scattering by vertical barriers}, Journal of Fluid Mechanics 294 (1995) 155--180.
\newblock \href {https://doi.org/10.1017/S0022112095002849} {\path{doi:10.1017/S0022112095002849}}.

\bibitem{McIver1985}
P.~McIver, {Scattering of Water Waves by Two Surface-piercing Vertical Barriers}, IMA Journal of Applied Mathematics 35~(3) (1985) 339--355.
\newblock \href {https://doi.org/10.1093/imamat/35.3.339} {\path{doi:10.1093/imamat/35.3.339}}.

\bibitem{porter1999rayleigh}
R.~Porter, D.~V. Evans, Rayleigh--bloch surface waves along periodic gratings and their connection with trapped modes in waveguides, Journal of Fluid Mechanics 386 (1999) 233--258.
\newblock \href {https://doi.org/10.1017/S0022112099004425} {\path{doi:10.1017/S0022112099004425}}.

\bibitem{callan_linton_evans_1991}
M.~Callan, C.~M. Linton, D.~V. Evans, Trapped modes in two-dimensional waveguides, Journal of Fluid Mechanics 229 (1991) 51–64.
\newblock \href {https://doi.org/10.1017/S0022112091002938} {\path{doi:10.1017/S0022112091002938}}.

\bibitem{maniar1997wave}
H.~D. Maniar, J.~N. Newman, Wave diffraction by a long array of cylinders, Journal of fluid mechanics 339 (1997) 309--330.
\newblock \href {https://doi.org/10.1017/S0022112097005296} {\path{doi:10.1017/S0022112097005296}}.

\bibitem{evans1987resonant}
D.~V. Evans, P.~McIver, Resonant frequencies in a container with a vertical baffle, Journal of Fluid Mechanics 175 (1987) 295--307.
\newblock \href {https://doi.org/10.1017/S0022112087000399} {\path{doi:10.1017/S0022112087000399}}.

\bibitem{Linton2001}
C.~M. Linton, P.~McIver, {Handbook of mathematical techniques for wave/structure interactions}, Chapman {\&} Hall, 2001.
\newblock \href {https://doi.org/10.1201/9781420036060} {\path{doi:10.1201/9781420036060}}.

\bibitem{Mei2005}
C.~C. Mei, M.~Stiassnie, D.~K.-P. Yue, {Theory and applications of ocean surface waves Part 1: Linear aspects}, World Scientific, 2005.
\newblock \href {https://doi.org/10.1142/5566} {\path{doi:10.1142/5566}}.

\bibitem{Mosig2018}
J.~E.~M. Mosig, {Contemporary wave-ice interaction models}, Ph.D. thesis, University of Otago (2018).

\bibitem{utsunomiya_taylor_1999}
T.~Utsunomiya, R.~Eatock~Taylor, Trapped modes around a row of circular cylinders in a channel, Journal of Fluid Mechanics 386 (1999) 259–279.
\newblock \href {https://doi.org/10.1017/S0022112099004437} {\path{doi:10.1017/S0022112099004437}}.

\bibitem{ooi2008analysis}
A.~Ooi, A.~Nikolovska, R.~Manasseh, Analysis of time delay effects on a linear bubble chain system, The Journal of the Acoustical Society of America 124~(2) (2008) 815--826.
\newblock \href {https://doi.org/10.1121/1.2945156} {\path{doi:10.1121/1.2945156}}.

\bibitem{WOLGAMOT2017232}
H.~A. Wolgamot, M.~H. Meylan, C.~D. Reid, Multiply heaving bodies in the time-domain: Symmetry and complex resonances, Journal of Fluids and Structures 69 (2017) 232--251.
\newblock \href {https://doi.org/10.1016/j.jfluidstructs.2016.11.012} {\path{doi:10.1016/j.jfluidstructs.2016.11.012}}.

\bibitem{chowdhury2023coupled}
S.~D. Chowdhury, L.~G. Bennetts, R.~Manasseh, A coupled damped harmonic oscillator model for arbitrary arrays of floating cylinders using homotopy methods, Physics of Fluids 35~(10) (2023).
\newblock \href {https://doi.org/10.1063/5.0165305} {\path{doi:10.1063/5.0165305}}.

\bibitem{bennetts2021complex}
L.~G. Bennetts, M.~H. Meylan, Complex resonant ice shelf vibrations, SIAM Journal on Applied Mathematics 81~(4) (2021) 1483--1502.
\newblock \href {https://doi.org/10.1137/20M1385172} {\path{doi:10.1137/20M1385172}}.

\bibitem{wilks2023_2d_time_domain}
B.~Wilks, M.~H. Meylan, F.~Montiel, S.~Wakes, Generalised eigenfunction expansion and singularity expansion methods for two-dimensional acoustic time-domain wave scattering problems (11 2023).
\newblock \href {https://doi.org/10.48550/arXiv.2311.11524} {\path{doi:10.48550/arXiv.2311.11524}}.

\bibitem{Ko1988}
D.~Y.~K. Ko, J.~R. Sambles, {Scattering matrix method for propagation of radiation in stratified media: attenuated total reflection studies of liquid crystals}, Journal of the Optical Society of America A 5~(11) (1988) 1863.
\newblock \href {https://doi.org/10.1364/josaa.5.001863} {\path{doi:10.1364/josaa.5.001863}}.

\bibitem{Bennetts2009}
L.~G. Bennetts, V.~A. Squire, {Wave scattering by multiple rows of circular ice floes}, Journal of Fluid Mechanics 639 (2009) 213--238.
\newblock \href {https://doi.org/10.1017/S0022112009991017} {\path{doi:10.1017/S0022112009991017}}.

\bibitem{sahoo2000trapping}
T.~Sahoo, M.~Lee, A.~Chwang, Trapping and generation of waves by vertical porous structures, Journal of engineering mechanics 126~(10) (2000) 1074--1082.
\newblock \href {https://doi.org/10.1061/(ASCE)0733-9399(2000)126:10(1074)} {\path{doi:10.1061/(ASCE)0733-9399(2000)126:10(1074)}}.

\bibitem{pagneux2013trapped}
V.~Pagneux, Trapped modes and edge resonances in acoustics and elasticity, in: R.~V. Craster, J.~Kaplunov (Eds.), Dynamic Localization Phenomena in Elasticity, Acoustics and Electromagnetism, Springer, 2013, pp. 181--223.

\bibitem{taylor2007modelling}
R.~Eatock~Taylor, On modelling the diffraction of water waves, Ship Technology Research 54~(2) (2007) 54--80.
\newblock \href {https://doi.org/10.1179/str.2007.54.2.002} {\path{doi:10.1179/str.2007.54.2.002}}.

\end{thebibliography}

\end{document}